\documentclass[extra] {gji}

\usepackage{xcolor}
\usepackage[colorlinks=True,%
            urlcolor=blue,%
            linkcolor=blue,%
            citecolor= blue,%
            pdfauthor={O. Barrois},%
            pdftitle={},%
            pdftex]{hyperref}
            
\usepackage{timet}
\usepackage{graphicx}
\usepackage{amsmath,amsfonts,amssymb}
\usepackage{longtable}
\usepackage{caption}
\usepackage{bm}

\renewcommand{\cite}{\citet}

\newcommand{\hlit}[1]{{\color{red} #1}}

\title[Propagation of QG MC waves over a geophysical background magnetic field]
	{Influence of simple and complex non-axisymmetric background magnetic fields on the propagation of QG MC waves}
\author[Barrois et al.]
  {O. Barrois$^1$, J. Aubert$^1$ \\
  $^1$ Universit\'e de Paris, Institut de Physique du Globe de Paris, CNRS, F-75005 Paris, France.}
\date{Received May 2025}
\pagerange{\pageref{firstpage}--\pageref{lastpage}}
\volume{XXX}
\pubyear{XXXX}

\begin{document}

\label{firstpage}

\maketitle

\begin{summary}

Magneto-Coriolis (QG-MC) waves are considered an important part of the rapid dynamics of the Earth's outer core.
The detailed characteristics of these waves are however still under scrutiny.
In this study we explore the sensibility of the QG-MC waves to the background magnetic field over which they propagate and to the frequency of the perturbation that can generate them in the Earth core.
We retrieve QG-MC waves propagating over a realistic background magnetic field by analysing the velocity fields, where they are most easily observed.
Concentrations of QG-MC waves in the magnetic field at the core surface in our model are reminiscent of recently observed geomagnetic jerks.
The QG-MC waves are weakly sensitive to the details of the background magnetic field during their travel in the bulk and their frequency at the core surface remains close to that of the initial perturbation.
This is a potential asset for the prediction of their evolution.
Moreover, the waves in the system exhibit a complex relation with the initial perturbation: when the frequency of the initial pulsation is greater than a threshold -- depending on the Alfvén speed and the geometric complexity of the medium -- inward QG-Alfvén waves are recovered at the core mantle boundary instead of QG-MC waves, and we find a continuum of waves from QG-MC to 
QG-Alfvén waves depending on the input frequency.
Thus, gradually increasing the input frequency in the system, we retrieve the dispersion relation for QG-MC waves with an evolution from a $k_s^4$ slope to a $k_s^1$ slope, where $k_s$ is the cylindrical radial wavenumber, as waves transition from QG-MC to QG-Alfvén waves.
Applying our results to the Earth's core, we expect to be able to recover QG-MC waves with confidence in the Earth core with periods between $57~y$ and $2.8~y$.

\end{summary}


\section{Introduction}
\label{sec:state_of_art}

The rapid dynamics of the geomagnetic field -- from inter-annual up to decadal timescales -- is driven by hydromagnetic waves, including Quasi-Geostrophic Magneto-Coriolis (QG-MC) waves which have been recently reconsidered as instrumental at these timescales.
QG-MC waves with different periods have been observed in the velocity time series inferred from the geomagnetic signals \citep{gillet2022satellite,istas2023transient} after having been discovered at inter-annual periods \citep{gerick2021fast}.
Their timescales are much shorter than the convection, or overturn, timescale of the core $\tau_u = d/U_c \approx 125\,y$ -- with $U_c \approx 18\,km/y$ a typical convective velocity of the fluid core, and with $d = 2260\,km$ the thickness of the Earth's core -- and much longer than the rotation timescale of the core $2 \pi\,\tau_\Omega = 2 \pi / \Omega = 1\,\mathrm{day}$ -- with $\Omega$ the rotation period of the Earth.
Their fundamental timescale is the Alfvén time of the core $\tau_{\cal A} = d\,\sqrt{\rho\,\mu}/B_\oplus$ -- where $\rho \simeq 10^4\,kg/m^3$ and $\mu = 4\pi \times 10 ^{-7}\,H/m$ are the density and the magnetic permeability of the fluid and $B_\oplus$ is the magnetic field of the Earth rms value, respectively -- and has an estimated value of $\tau_{\cal A} \simeq 2\,y$ for the Earth outer core obtained from an rms field strength of $B_\oplus \sim 4\,mT$ \citep{gillet2010fast}.

In the recent years, the improvement of the geomagnetic field models \citep[{\it e.g.},][]{lesur2022rapid,finlay2020chaos}, and the use of data assimilation techniques to predict the evolution of the geomagnetic field \citep{barrois2018assimilation,aubert2020recent,istas2023transient} has produced satisfactory results to explain the signals recovered at the core mantle boundary (CMB).
But the fundamental question about the possibility to improve our predictions compared to a simple extrapolation \citep{alken2021international} remains.
Using hydrodynamics waves to probe the Earth's core interior is not new \citep{buffett2019equatorially,gillet2015planetary} and some waves like the torsional Alfvén waves have given good insights on both the length-of-day and the core surface flows dynamics \citep{gillet2010fast} even if these torsional Alfvén waves do not account for all of the inter-annual geomagnetic signals \citep{chulliat2014geomagnetic}.
More generally, the study of a variety of periodic signals found in the geophysical records have provided insightful knowledge on the otherwise directly inaccessible deep layers of the Earth \citep{rekier2022earth,triana2022core,rosat2023intradecadal,schwaiger2024wave,cazenave2025earth}.
And with the recently discovered QG-MC waves of suitable periods which could be sensible to the long term velocity and 
magnetic fields \citep[{\it e.g.},][]{gerick2024interannual}, hopes are it would potentially unlock a new level of understanding on the deep structure of the Earth's core and allow for an improvement of our predictions \citep{gillet2022satellite}.
This question is directly relevant in our present study and is one of the main motivations underlying this work.

Our approach builds upon 3D model computations \cite[{\it e.g.},][]{aubert2023state} in a reduced framework taking advantage of the timescales separation between the slow secular processes and the fast inter-annual dynamics of the Earth core, which is a common strategy for studying axisymmetric or non-axisymmetric perturbations and characterising the rapid dynamics of the geodynamo system \citep{jault2008axial,gillet2011rationale,gerick2021fast}.
Considering that the slower convective dynamics is static compared to the inter-annual waves, we adopt a perturbation approach and linearise the magneto-hydro-dynamics equations around a stationary background state which ensures the timescales separation of the different processes.
This is a continuation of our previous work \citep{barrois2024characterization} as we add more complexity by using a more geophysical background magnetic field from the path model dynamo models \citep{aubert2017spherical}.
And we also study the response of the system to a monochromatic periodic forcing instead of its impulsional response.

The paper is organised as follows.
We briefly describe the context and the geophysical background field used in our study in Section~\ref{sec:Methods}, then present our results in Section~\ref{sec:Results}, before discussing and analysing the conclusions of this work in Section~\ref{sec:Conclusion}.
More details are given about the methodology of our study in the Appendix~\ref{sec:Append-A-Methods}.

\section{Methods}
\label{sec:Methods}

\subsection{System solved}
\label{sec:Equations}

In this study, we follow \cite{jault2008axial,barrois2024characterization} and use a reduced system of equations (see Methods details~\ref{sec:MHD-lin_equations}, Eq.~\ref{eq:momentum_noT_linearised}-\ref{eq:induction_linearised}) linearised around an arbitrary background velocity field ${\bf U}_0$ and a chosen background magnetic field ${\bf B}_0$ (see Fig.~\ref{fig:B_background_PB}).
Our model describes the dynamics of a rapidly rotating thick spherical shell filled with a conducting incompressible fluid and solves for the velocity field ${\bf U}$ and the magnetic field ${\bf B}$ of this fluid.
More precisely, it solves for the velocity perturbation ${\bm u}$ and magnetic perturbation ${\bf b}$ fields, {\it i.e.} ${\bf U} \equiv {\bf U}_0 + {\bm u}$ and ${\bf B} \equiv {\bf B}_0 + {\bf b}$ and we make use of the spherical coordinates system $(r, \theta, \phi)$ with unit vectors $({\bm e}_r, {\bm e}_\theta, {\bm e}_\phi)$, even though the cylindrical coordinates system $(s, \phi, z)$ with unit vectors $({\bm e}_s, {\bm e}_\phi, {\bm e}_z)$ might alternatively be used.
The thick shell is assumed to have a thickness $d = r_o - r_i$ with an aspect ratio $r_i/r_o = 0.35$ -- with $r_o = 3,485\,km$ and $r_i= 1,225\,km$ respectively the outer and inner core surface radii -- rotating at a constant angular velocity aligned with the rotation axis ${\bm \Omega} = \Omega {\bm e}_z$, which reflects the outer core of the Earth.
A solid inner core is included in the system, which is able to rotate independently -- the full kinetic momentum will be conserved --, and which has the same electrical conductivity as that of the outer core, $\sigma \simeq 0.4\,S.m^{-1}$.
The boundary conditions are insulating at $r = r_o$ and conducting at $r = r_i$ for the magnetic field, and are stress-free at both $r_i$ and $r_o$ for the velocity field.

We focus here on a series of simulations conducted at moderate parameters aiming at reproducing the Earth core conditions and the observed geomagnetic waves, {\it i.e.} with a high rotation rate, a strong magnetic field and a low level of attenuation. 
More details about our methodology and the equations of our system can be found in the Appendix~\ref{sec:MHD-lin_equations}.

\subsection{Magnetic base state}
\label{sec:B0}

\begin{figure*}
\centering{
	\includegraphics[width=0.45\linewidth]{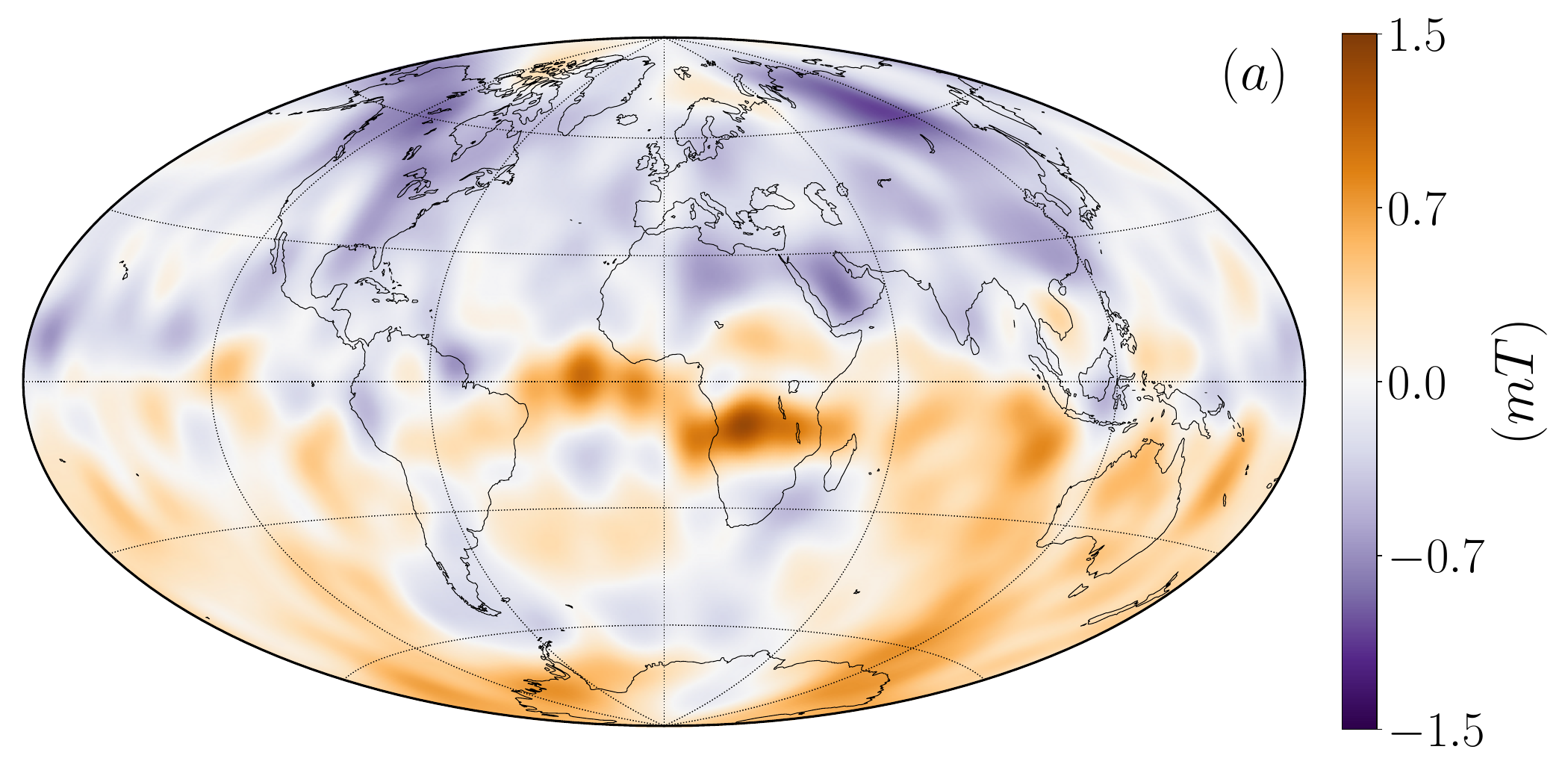}
    \includegraphics[width=0.17\linewidth]{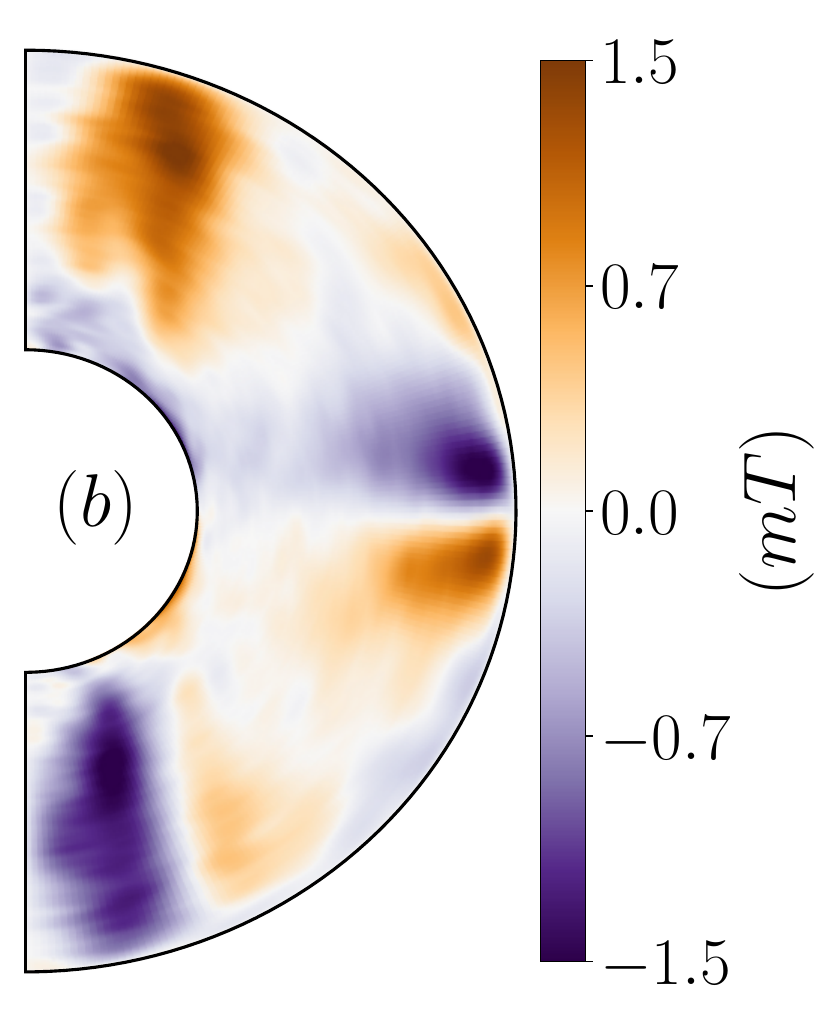}
    \includegraphics[width=0.27\linewidth]{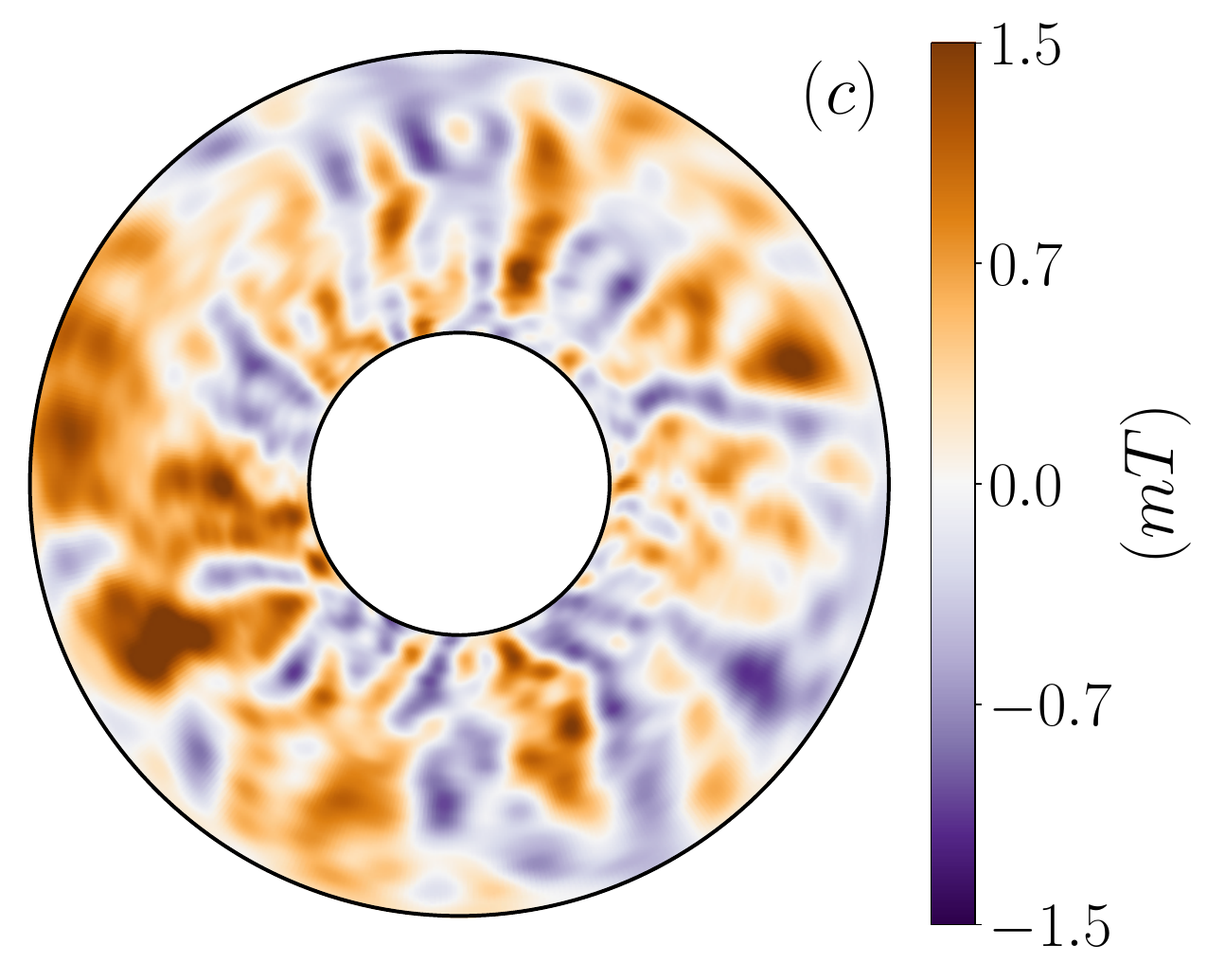}
}
	\caption{
	Complex background magnetic field ${\bf B}_0$ used in this study:
    (a) radial component of the magnetic field $B_{0,r}$ at the core surface,
    (b) axisymmetric azimuthal magnetic field $\dfrac{1}{2\pi} \int_{0}^{2\pi} B_{0,\phi}\,\mathrm{d}\phi$, 
    (c) and $B_{0,r}$ in the equatorial plane.
	}
	\label{fig:B_background_PB}
\end{figure*}

In order to best represent the magnetic structure of the Earth's core, we use a data assimilation based recent inference of the magnetic field inside the Earth's core.
Specifically, we use the ensemble mean of sequence A from \cite{aubert2023state} at epoch $2000$.
Figure~\ref{fig:B_background_PB} displays the main characteristics of this background field ${\bf B}_0$ that, by construction, matches the present geomagnetic field at the CMB (as can be seen in Fig.~\ref{fig:B_background_PB}~a) and retains the properties of the dynamo model in its interior (Fig.~\ref{fig:B_background_PB}~b-c).
This background state is highly heterogeneous and carries non-symmetric components, although it remains mostly dipolar.

Note that, as a reference, we also use a simple background magnetic field from \cite{barrois2024characterization} (see their Fig.~$1$).
Respectively, the cases computed using the simple background are labeled Cases S (or SB-$\#$ for more specific cases) and the cases computed using the complex background are labeled Cases C (or CB-$\#$ for more specific cases) in Table~\ref{tab:run_list}.

\subsection{Initial conditions}
\label{sec:Init}

At the start of our computations, the system is at rest with respect to the reference frame of rotation.
To set the system in motion, a non-axisymmetric monochromatic periodic forcing is imposed in the force balance just above $r_i$.
The periodic perturbation function ${\cal F}_p$ is set at a specific sectorial spherical harmonic, Gaussian in the radial direction and periodic in time, such that
\begin{align}
\label{eq:impulse_Force}
{\cal F}_p(r, \theta, \phi, t) = \dfrac{1}{\sqrt{2\pi}\,\sigma^*} e^{-\dfrac{1}{2}\left( \dfrac{r - r_i}{\sigma^*} \right)^2}  \, Y_3^3(\theta, \phi) \, \sin(\omega_i\,t)\,,
\end{align}
where the $Y_\ell^m(\theta, \phi)$ are the spherical harmonics into which both the velocity and magnetic fields are decomposed -- with $\ell$ and $m$ respectively the spherical harmonic degree and order of the decomposition --, $\sigma^* = 0.01$ controls the sharpness of the Gaussian, and $\omega_i$ is the input pulsation of the system which is a free parameters that controls the frequency of the perturbation -- reported in Table~\ref{tab:run_list} for our set of simulations (expressed in Alfvén times).
Note that the input pulsation is related to the input period with the relation $\omega_i = 2\pi / T_i$.

\section{Results}
\label{sec:Results}

The periodic perturbation is switched on at $t=0$ and we wait for a steady state to take place -- usually after a time in the simulations corresponding to a few years -- before analysing the results that are presented throughout this section.
Note that the input parameters and some additional observations for all our $22$ simulations are reported in Table~\ref{tab:run_list} (Appendix~\ref{sec:Append-B-Results}).

\subsection{Response to a periodic forcing}
\label{sec:Steady-state}

\begin{figure*}
\centering{
    \includegraphics[width=.31\linewidth]{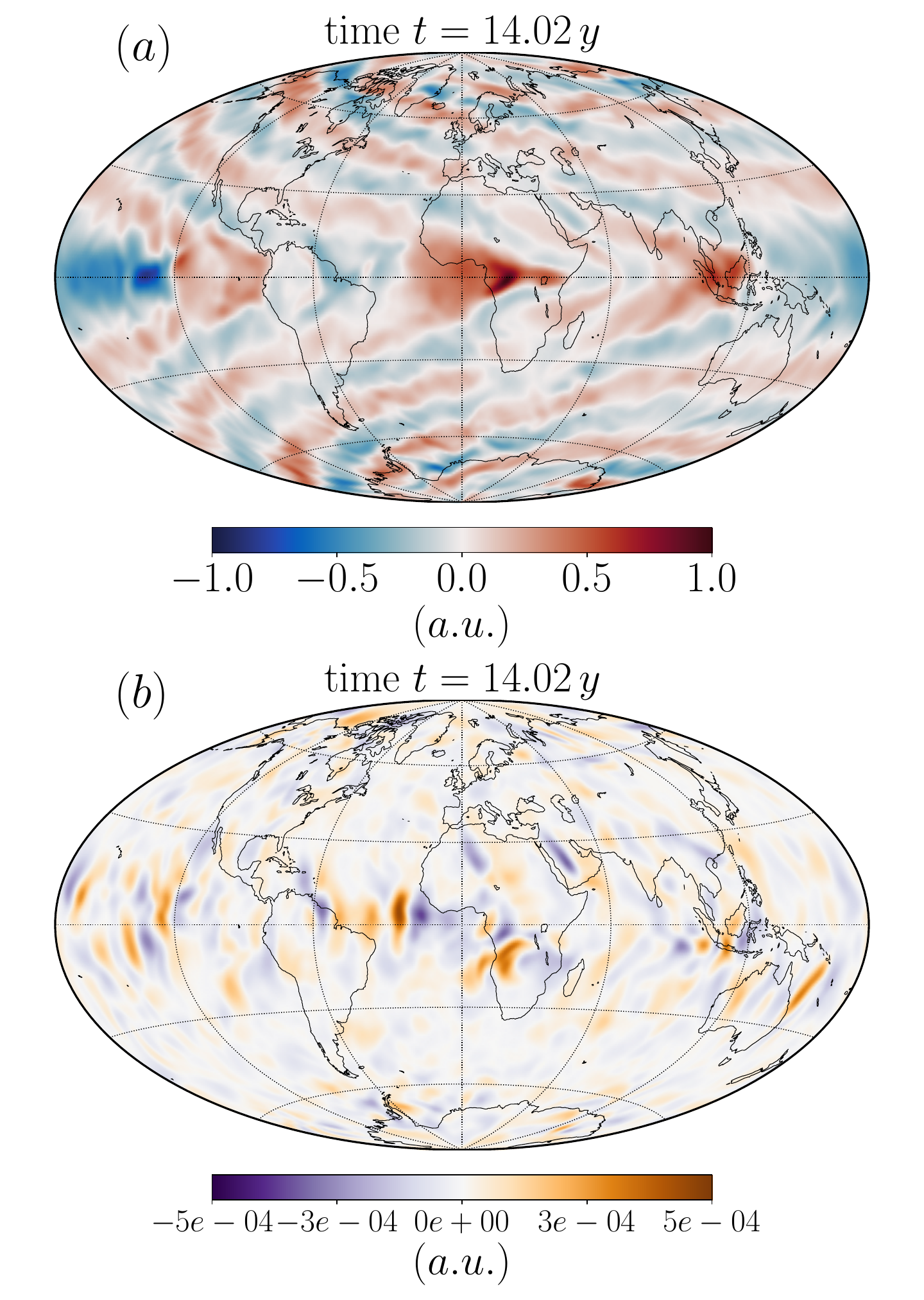}
    \includegraphics[width=.33\linewidth]{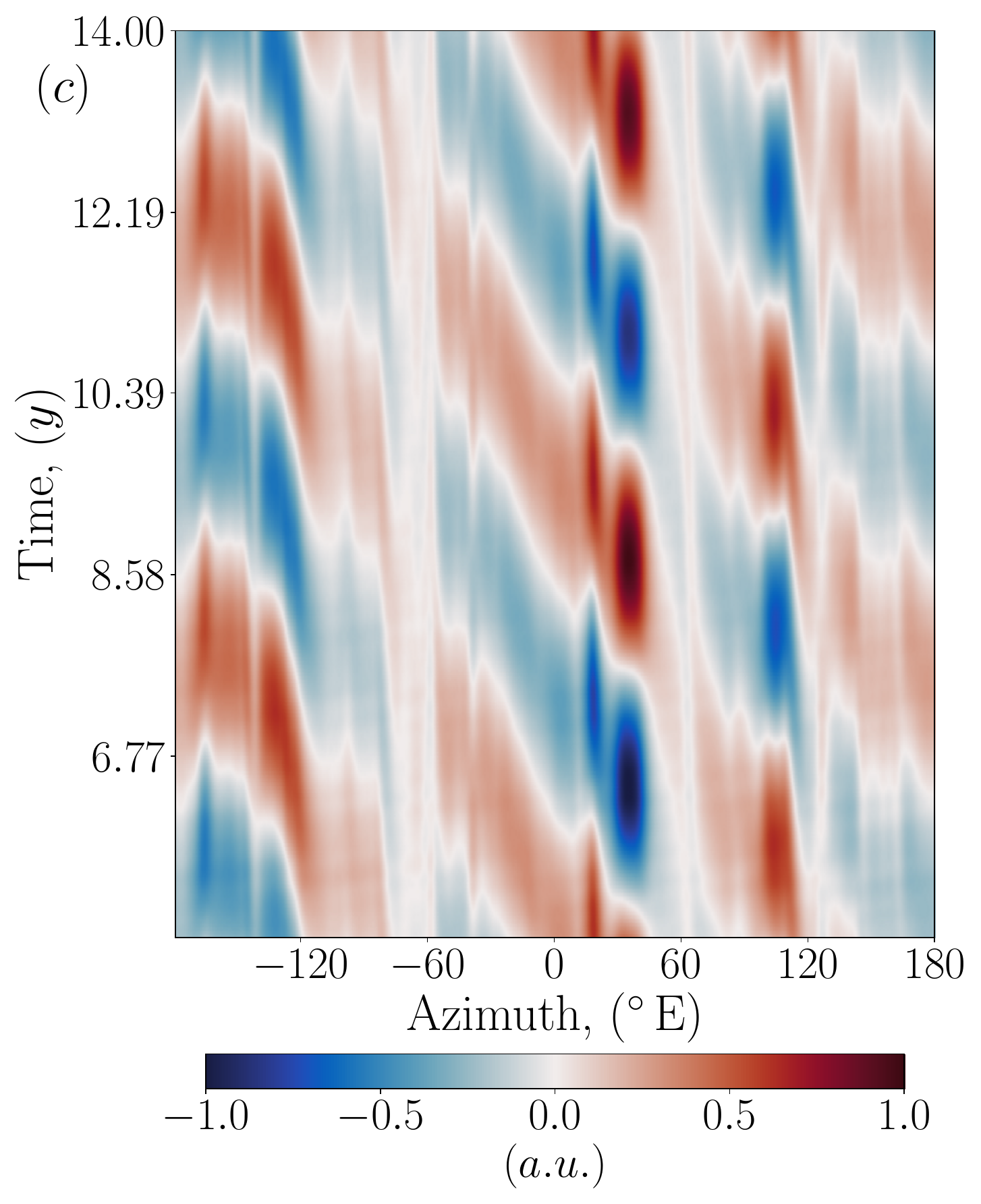}
	\includegraphics[width=.33\linewidth]{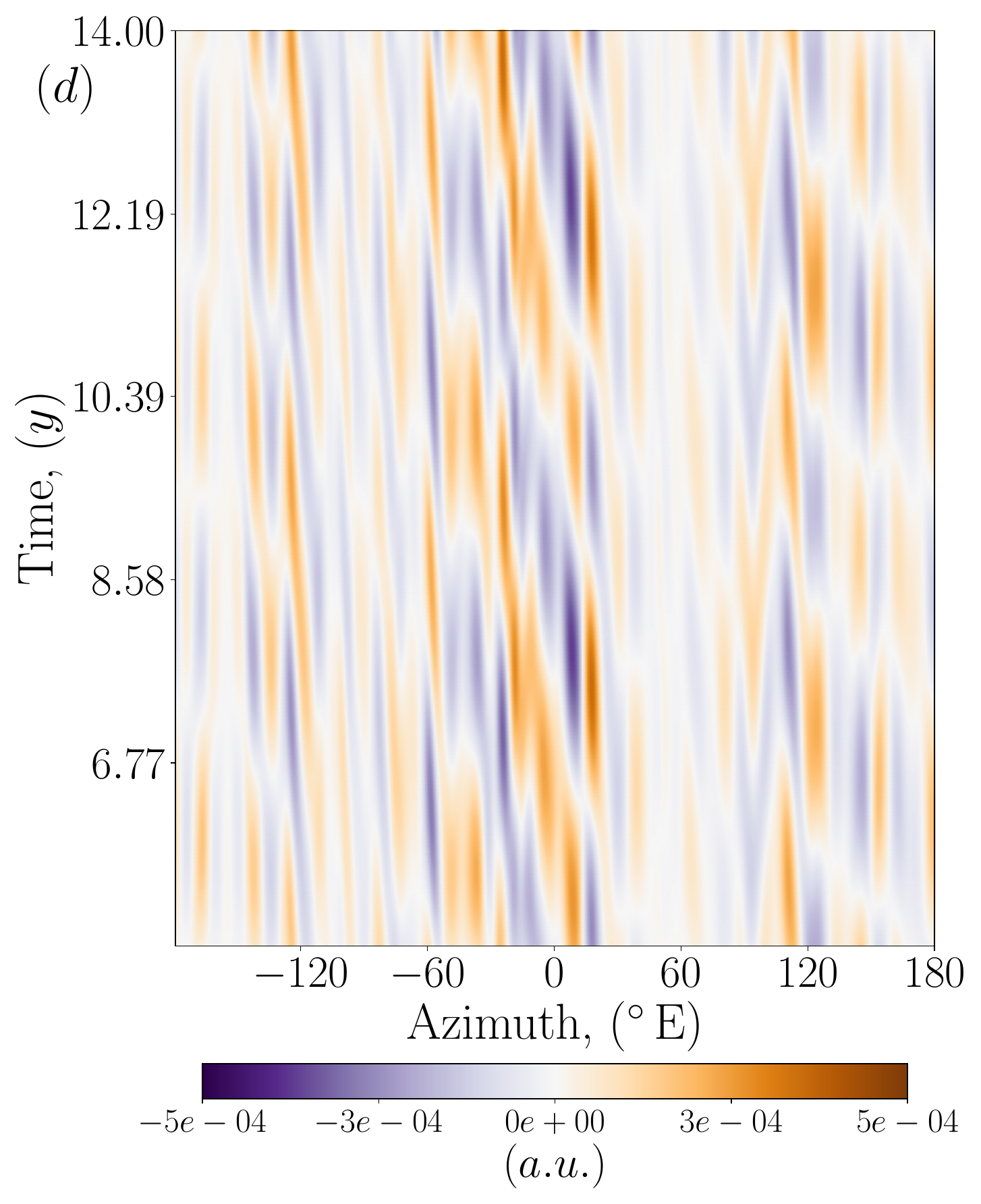}
    }
	\caption{
    Snapshots of the non-axisymmetric azimuthal velocity field at the outer boundary (a), of the radial perturbation magnetic field at the outer boundary (b), and time-azimuthal plots of the non-axisymmetric azimuthal velocity field at the equator (c), and of the radial perturbation magnetic field at the equator (d), for Case CB-1 using the complex magnetic background field and an input period corresponding to $T_i = 4.4\,y$.
    Note that the velocities have been normalised by their respective maximum value and that both the velocity and the magnetic fields have arbitrary units.
	}
	\label{fig:Maps-CMB_t_evolution_PB}
\end{figure*}

\begin{figure*}
\centering{
    \includegraphics[width=.31\linewidth]{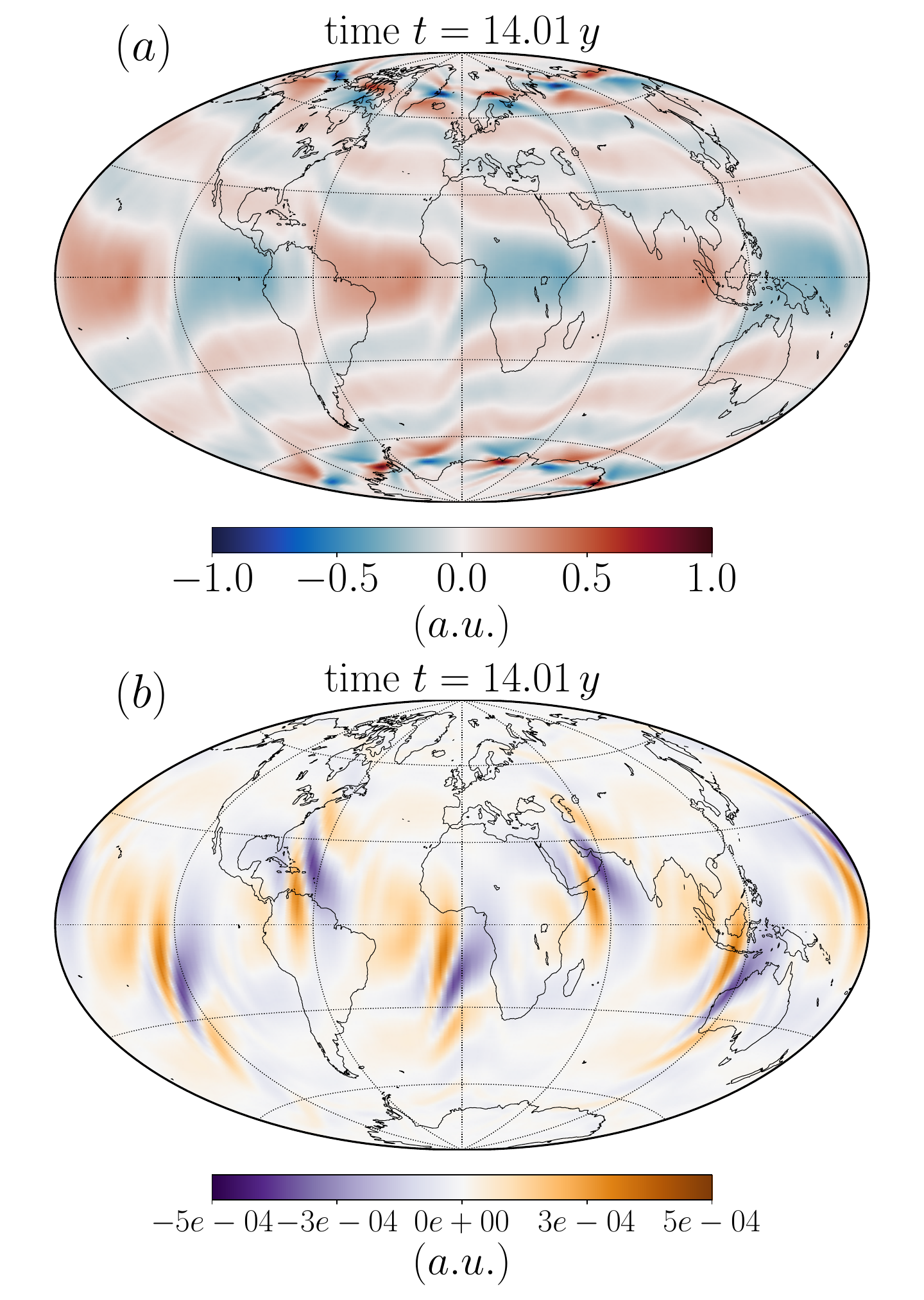}
    \includegraphics[width=.33\linewidth]{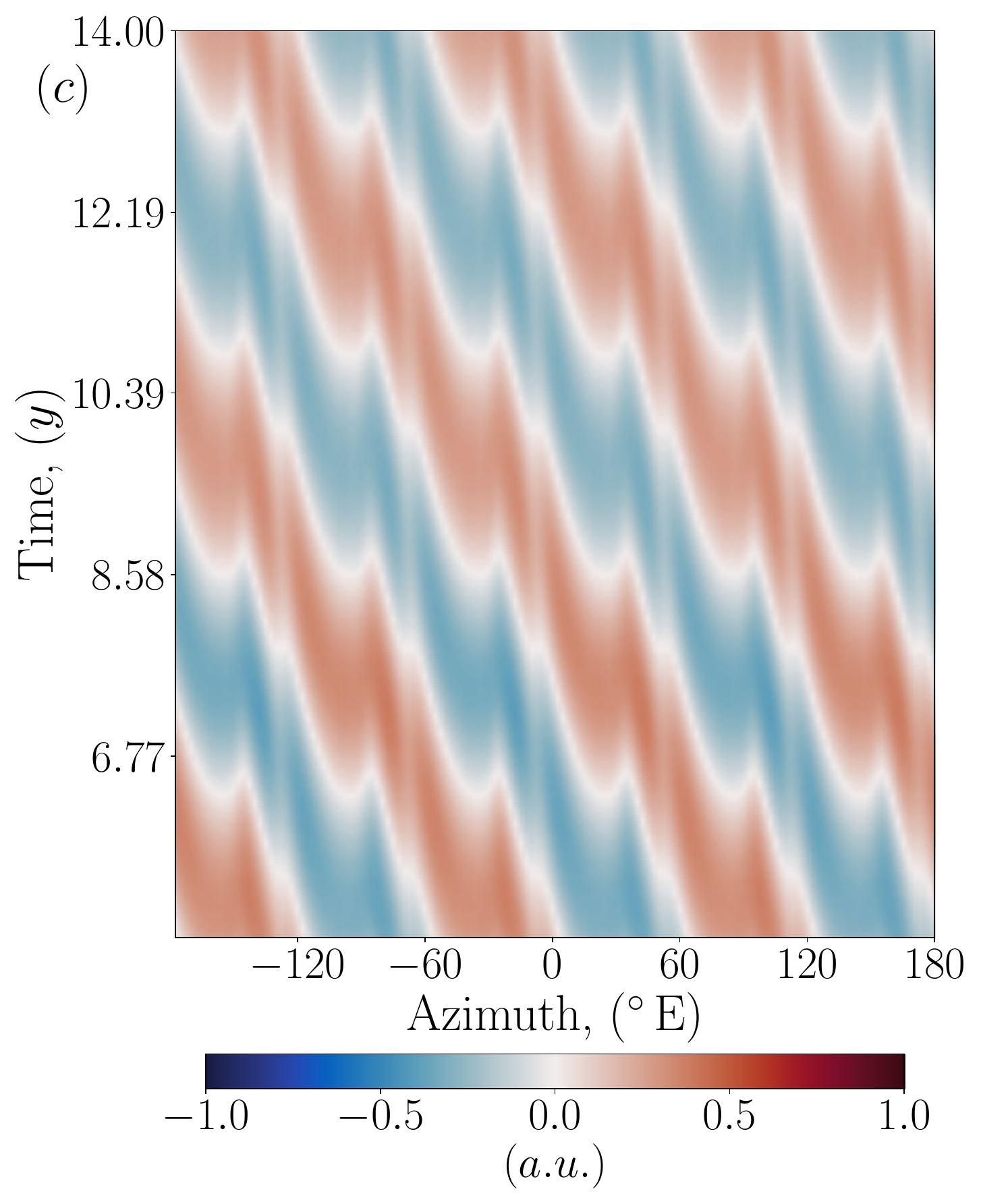}
	\includegraphics[width=.33\linewidth]{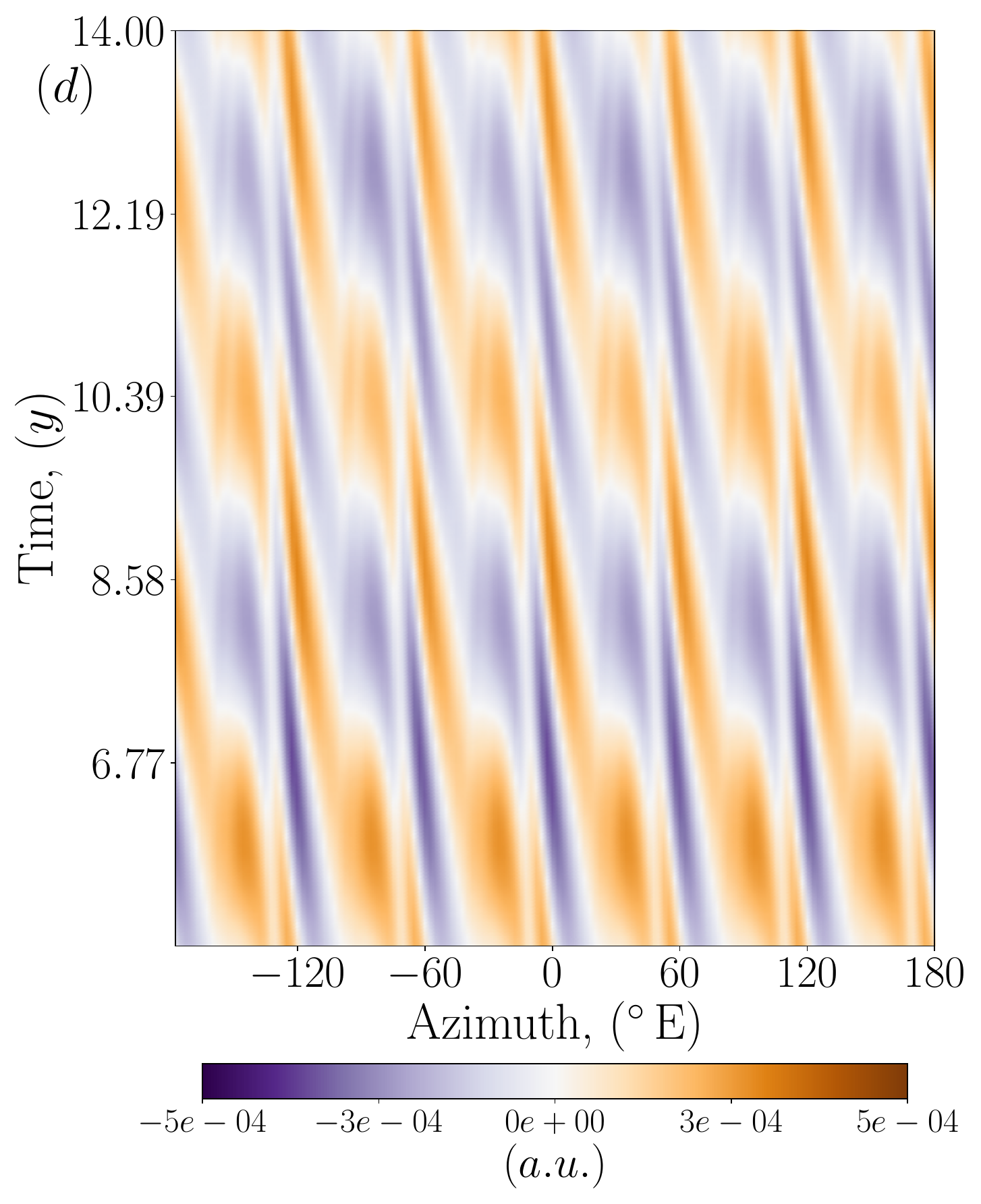}
    }
	\caption{
    Same as Figure~\ref{fig:Maps-CMB_t_evolution_PB} for Case SB-1 using a simple magnetic background field and an input period corresponding to $T_i = 4.4\,y$.
	}
	\label{fig:Maps-CMB_t_evolution}
\end{figure*}

\begin{figure*}
\centering{
    \includegraphics[width=.31\linewidth]{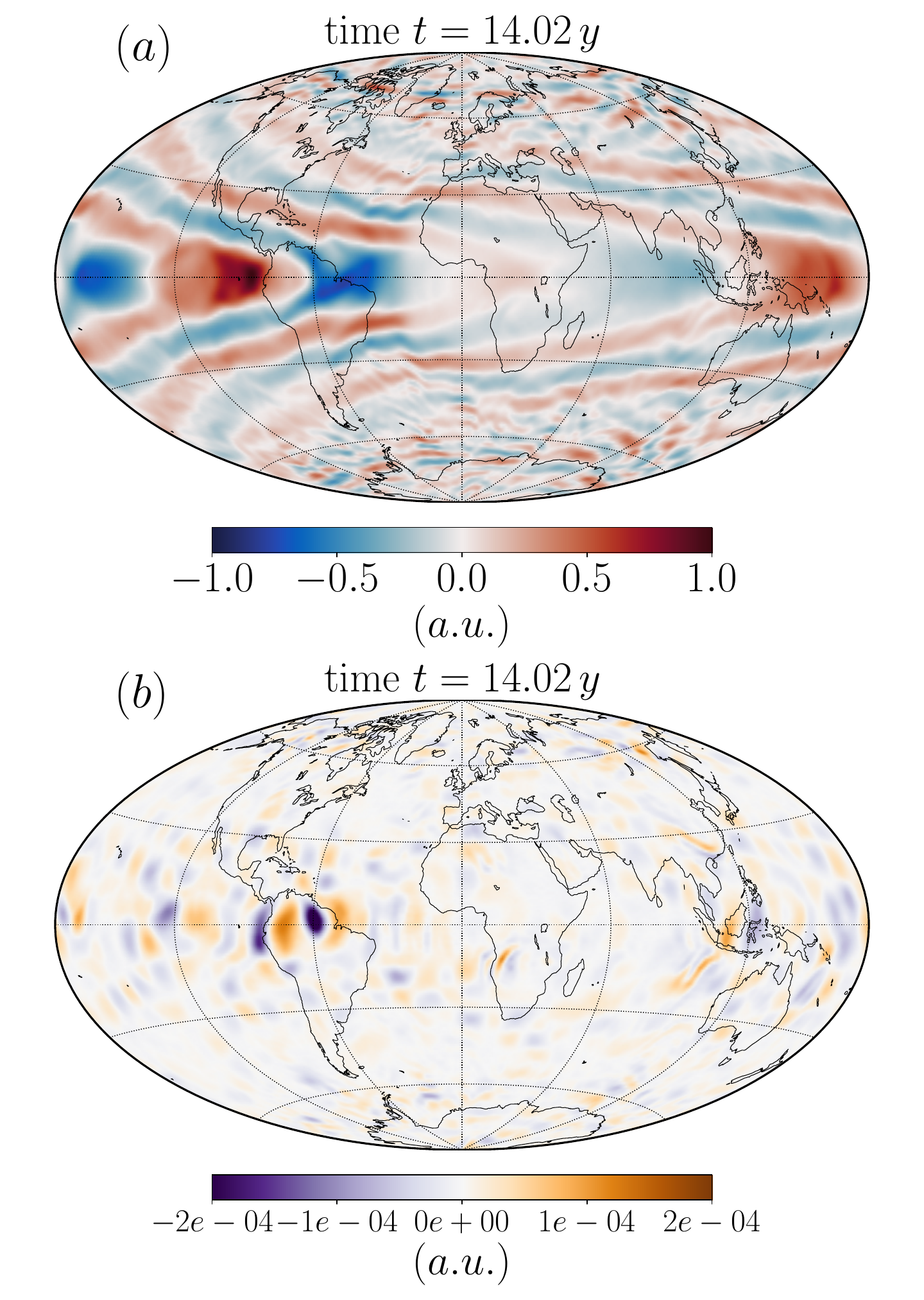}
    \includegraphics[width=.33\linewidth]{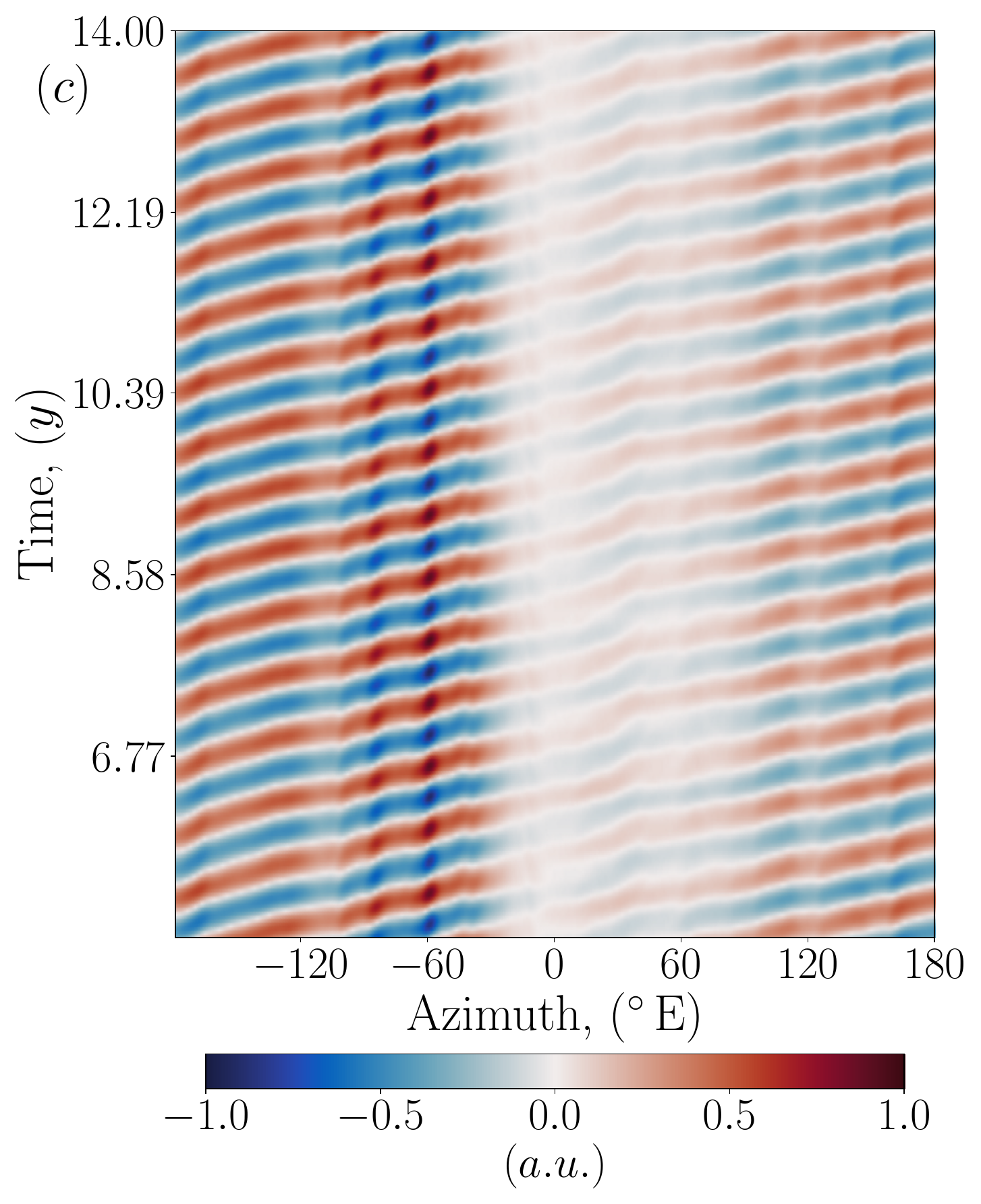}
	\includegraphics[width=.33\linewidth]{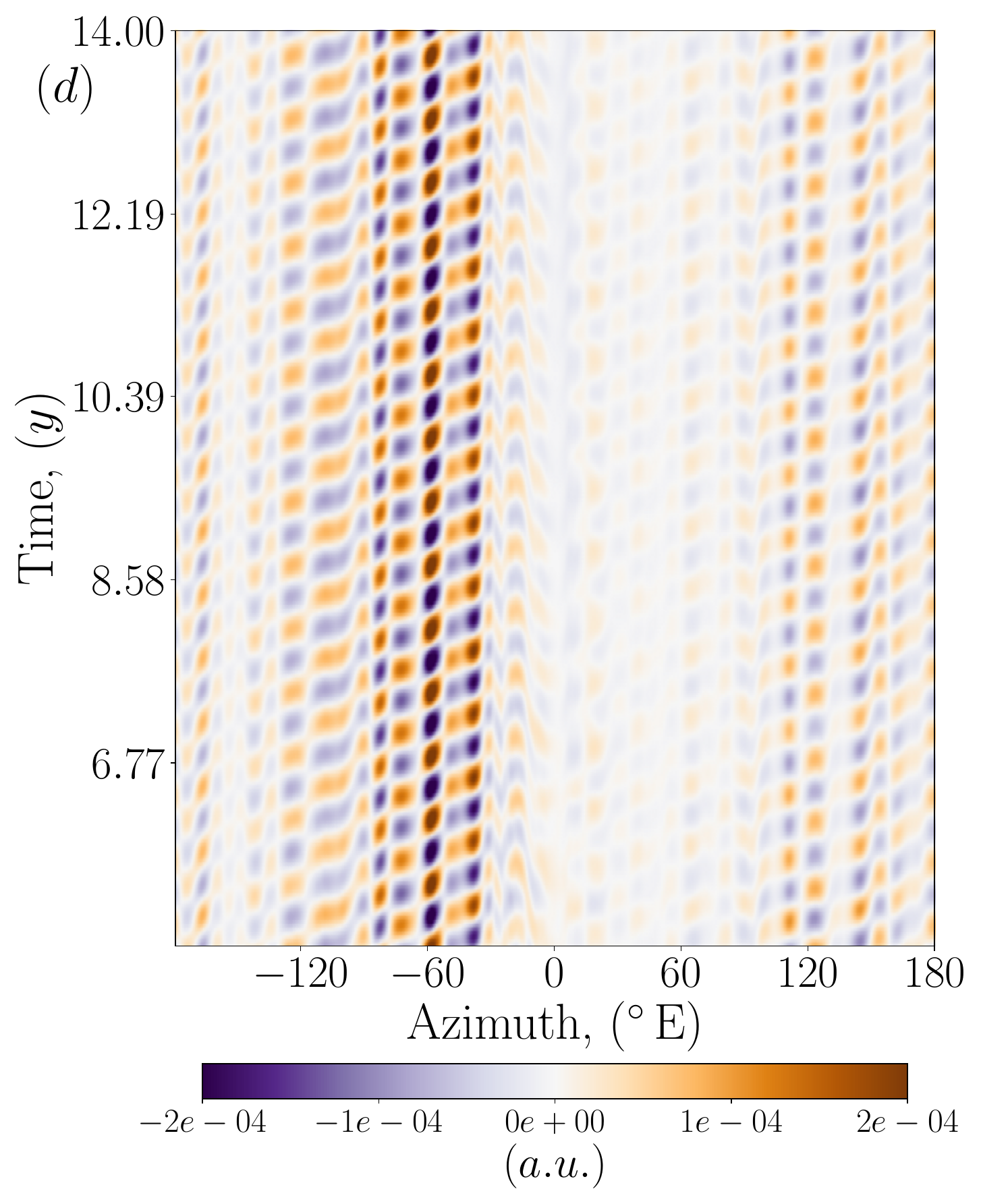}
    }
	\caption{
	Same as Figure~\ref{fig:Maps-CMB_t_evolution_PB} for Case CB-2 using the complex magnetic background field and an input period corresponding to $T_i = 0.6\,y$.
	}
	\label{fig:Maps-CMB_t_evolution_PB_faster}
\end{figure*}

For Case CB-1, using a forcing period corresponding to $T_i = 4.4\,y$ and the background magnetic field of Fig.~\ref{fig:B_background_PB}, snapshots of the azimuthal velocity at the core surface (Fig.~\ref{fig:Maps-CMB_t_evolution_PB}~a) reveal the wedged patterns that focus on the equator which are characteristic of QG-MC waves \citep{gillet2022satellite,gerick2024interannual}.
In the radial magnetic field (Fig.~\ref{fig:Maps-CMB_t_evolution_PB}~b) the QG-MC waves signature take the form of irregular patches mostly found at low latitudes and slightly elongated in the latitudinal direction.
In all panels of Fig.~\ref{fig:Maps-CMB_t_evolution_PB} we find concentration of waves in regions corresponding to that of under the Atlantic ($\approx 0 ^\circ\,$E), under the Pacific ($\approx -130 ^\circ\,$E) and under Indonesia ($\approx 120 ^\circ\,$E).
These patterns show a westward propagation and are mostly visible in the equatorial regions for both the azimuthal velocity and the radial magnetic fields, an observation that is very clear in the time-azimuth diagrams for both quantities (Fig.~\ref{fig:Maps-CMB_t_evolution_PB}~c-d).

The same observations can be made for Case SB-1, this time using the idealised background state introduced in \cite{barrois2024characterization} (Fig.~$1$) and an input period also corresponding to $T_i = 4.4\,y$. 
Despite the simplicity of this simpler background state, similar wedged patterns that focus at the equator (Fig.~\ref{fig:Maps-CMB_t_evolution}~a), the characteristic latitudinal shapes (Fig.~\ref{fig:Maps-CMB_t_evolution}~b) and the westward propagation (Fig.~\ref{fig:Maps-CMB_t_evolution}~c-d) can be retrieved again.
However, compared to the complex case, the patterns are perfectly regular, do not show regional heterogeneities and the features of the QG-MC waves appear more clearly, especially in the radial magnetic field where the QG-MC waves take the form of latitudinally elongated patches that converge toward the equator (Fig.~\ref{fig:Maps-CMB_t_evolution}~b).

Conversely, Case CB-2, using the complex magnetic field background and a faster input period corresponding to $T_i = 0.6\,y$ (Fig.~\ref{fig:Maps-CMB_t_evolution_PB_faster}), does not display the same characteristics as the two former cases.
The patterns at the CMB are restricted to a thinner equatorial band and are located in a region corresponding to that of under the South America (Fig.~\ref{fig:Maps-CMB_t_evolution_PB_faster}~a-b).
Moreover, we rather observe a slow eastward drift in both the velocity and the magnetic fields (Fig.~\ref{fig:Maps-CMB_t_evolution_PB_faster}~c-d).
Therefore, the waves recovered at the CMB using this faster input period are probably no longer QG-MC waves but likely QG-Alfvén waves -- considering the waves that can be generated in this reduced setup \citep{barrois2024characterization}.

Comparison of Fig.~\ref{fig:Maps-CMB_t_evolution_PB} and \ref{fig:Maps-CMB_t_evolution} indicates that QG-MC waves are weakly sensitive to the details of the background state, and remain clearly observable at the core surface even when using a complex background magnetic field.
It appears easier to observe the QG-MC waves in the velocity field, rather than in the magnetic field, in the more complex case -- a conclusion already drawn by \cite{gillet2022satellite}.
In addition, we can see that the QG-MC wave patterns in the radial magnetic field of the complex case (Fig.~\ref{fig:Maps-CMB_t_evolution_PB}~b-d) have distinct features in the Pacific and the Atlantic hemispheres, especially compared with the patterns of the simple case where the QG-MC waves look similar all over the core surface (Fig.~\ref{fig:Maps-CMB_t_evolution}).
For example, we can observe the concentration of waves below the Atlantic in the magnetic field of the complex case (Fig.~\ref{fig:Maps-CMB_t_evolution_PB}~b-and-d) similarly to what have been reported for some of the jerks of the satellite observation era -- like the $2007$ jerk which also focuses at these longitudes \citep{chulliat2010core,chulliat2015fast,aubert2019geomagnetic}.
These results suggest that the wave patterns carry information on the magnetic field at the CMB at least and that they might also carry information about the deep state of the magnetic field.
And from these observations, we can guess that the signal in the Pacific region may be more suitable for the observation of QG-MC waves in the geomagnetic time series due to the quieter geomagnetic activity in the Pacific hemisphere.

\subsection{Force balance}
\label{sec:col_bal}

\begin{figure*}
\centering{
    \includegraphics[width=0.31\linewidth]{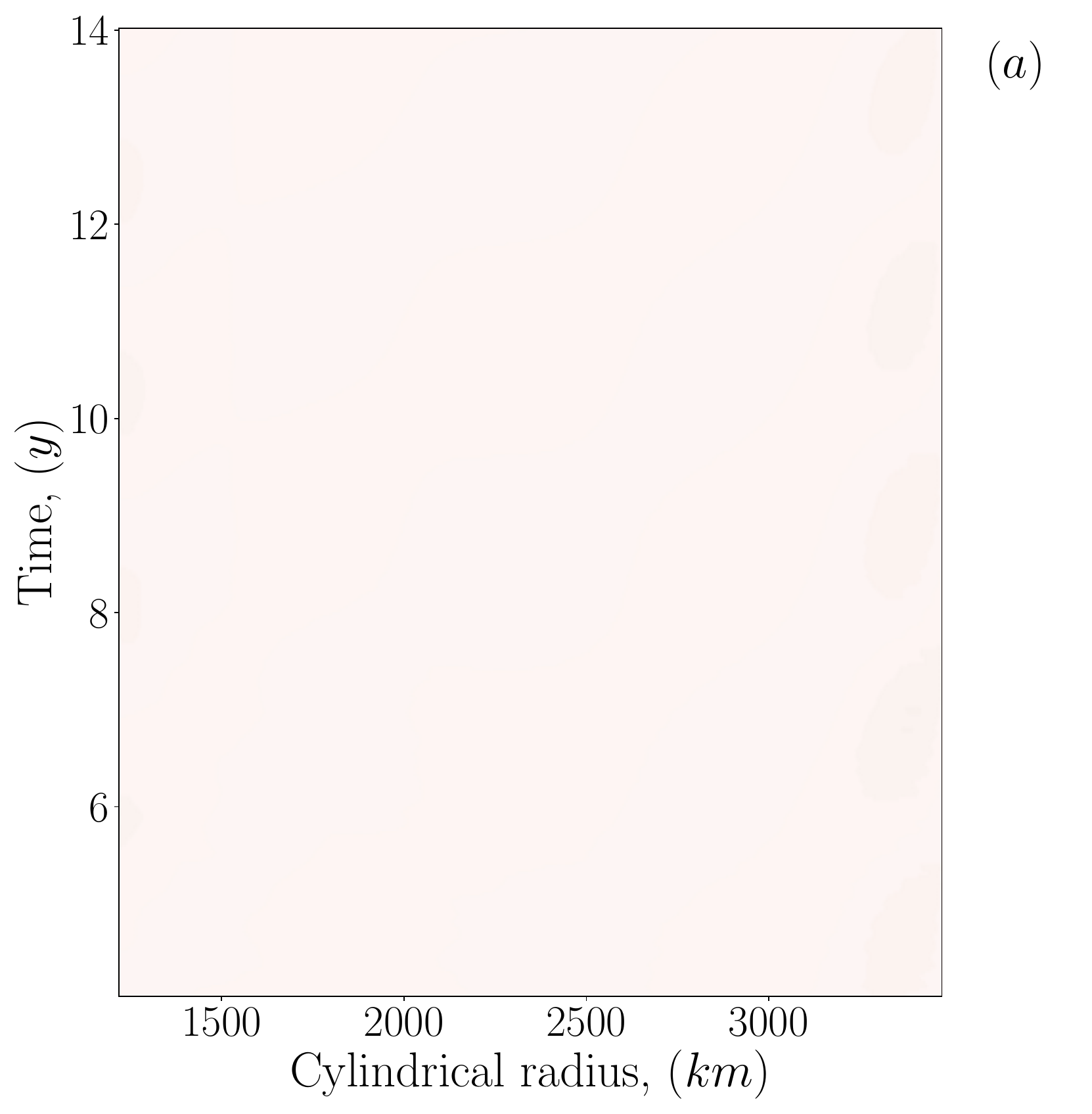}
    \includegraphics[width=0.31\linewidth]{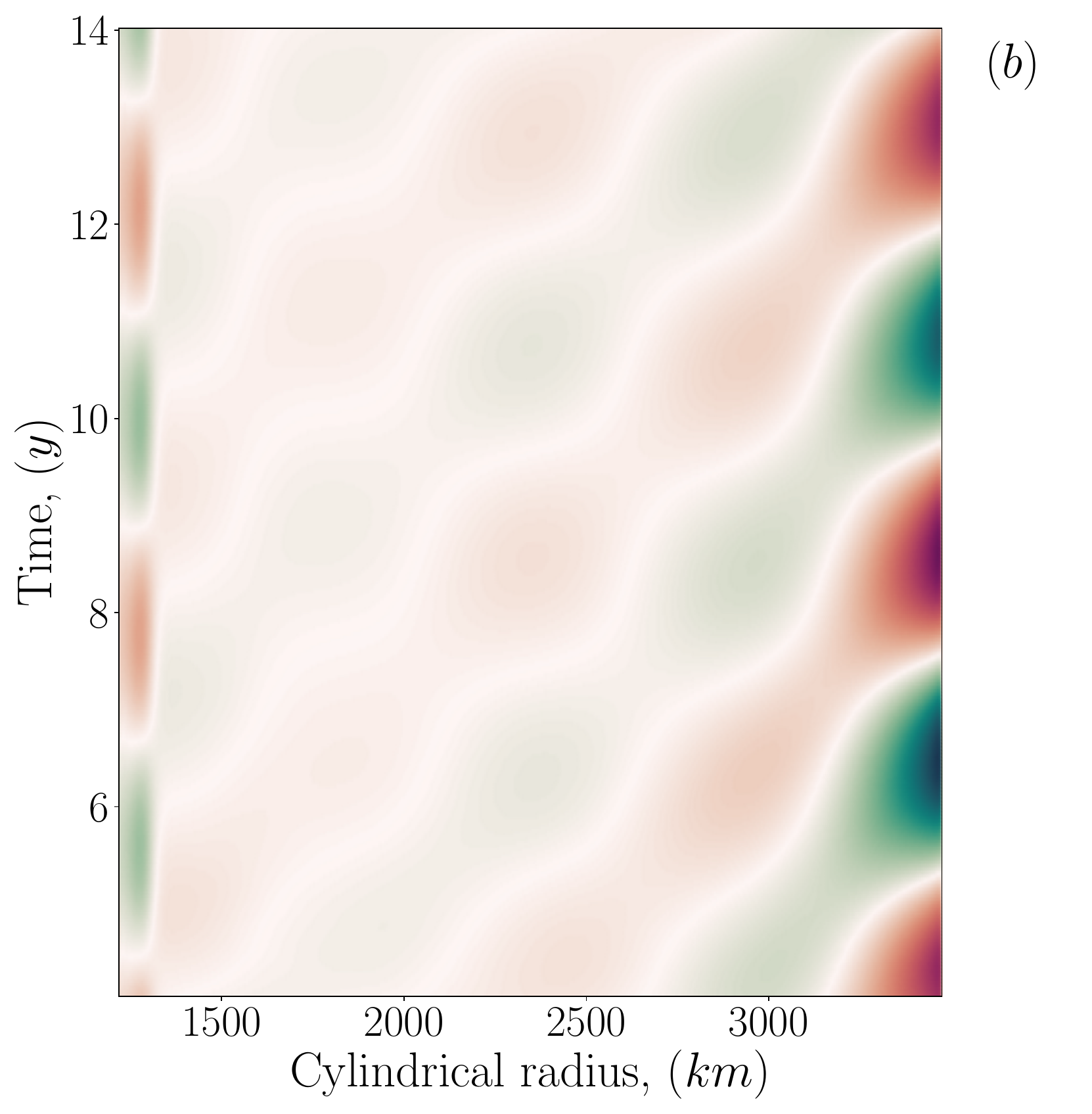}
	\includegraphics[width=0.33\linewidth]{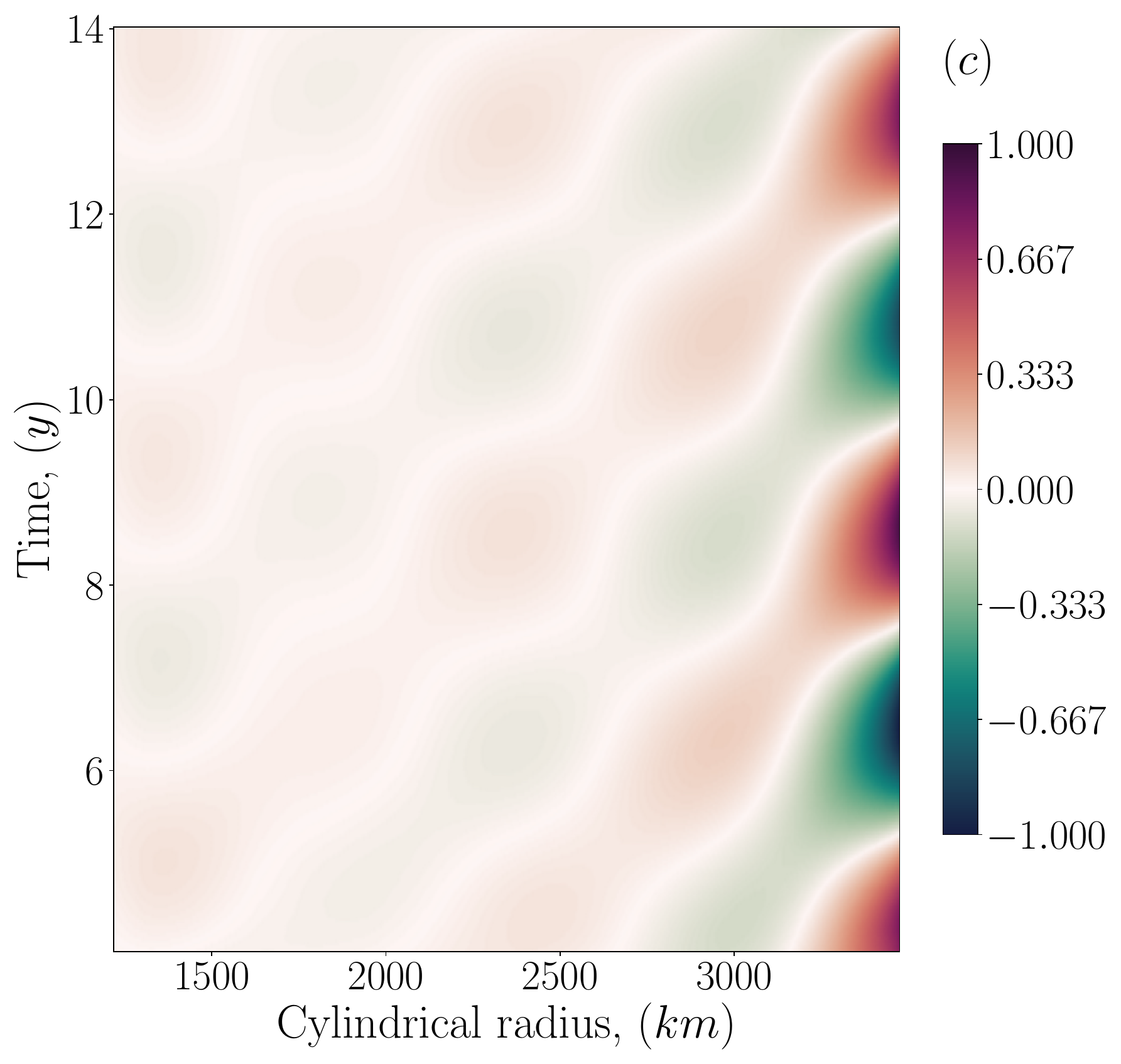}
}
\centering{
    \includegraphics[width=0.31\linewidth]{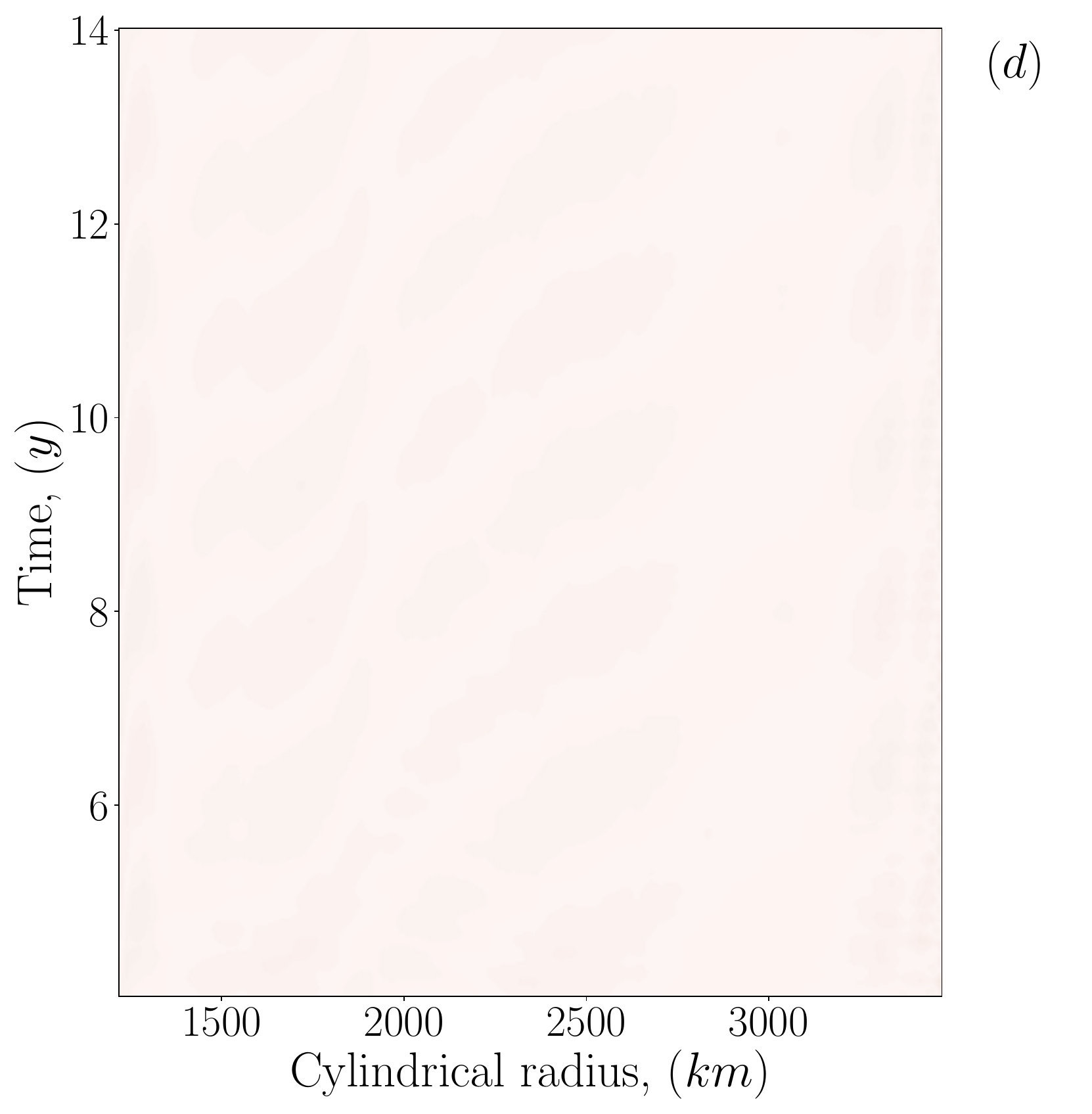}
    \includegraphics[width=0.31\linewidth]{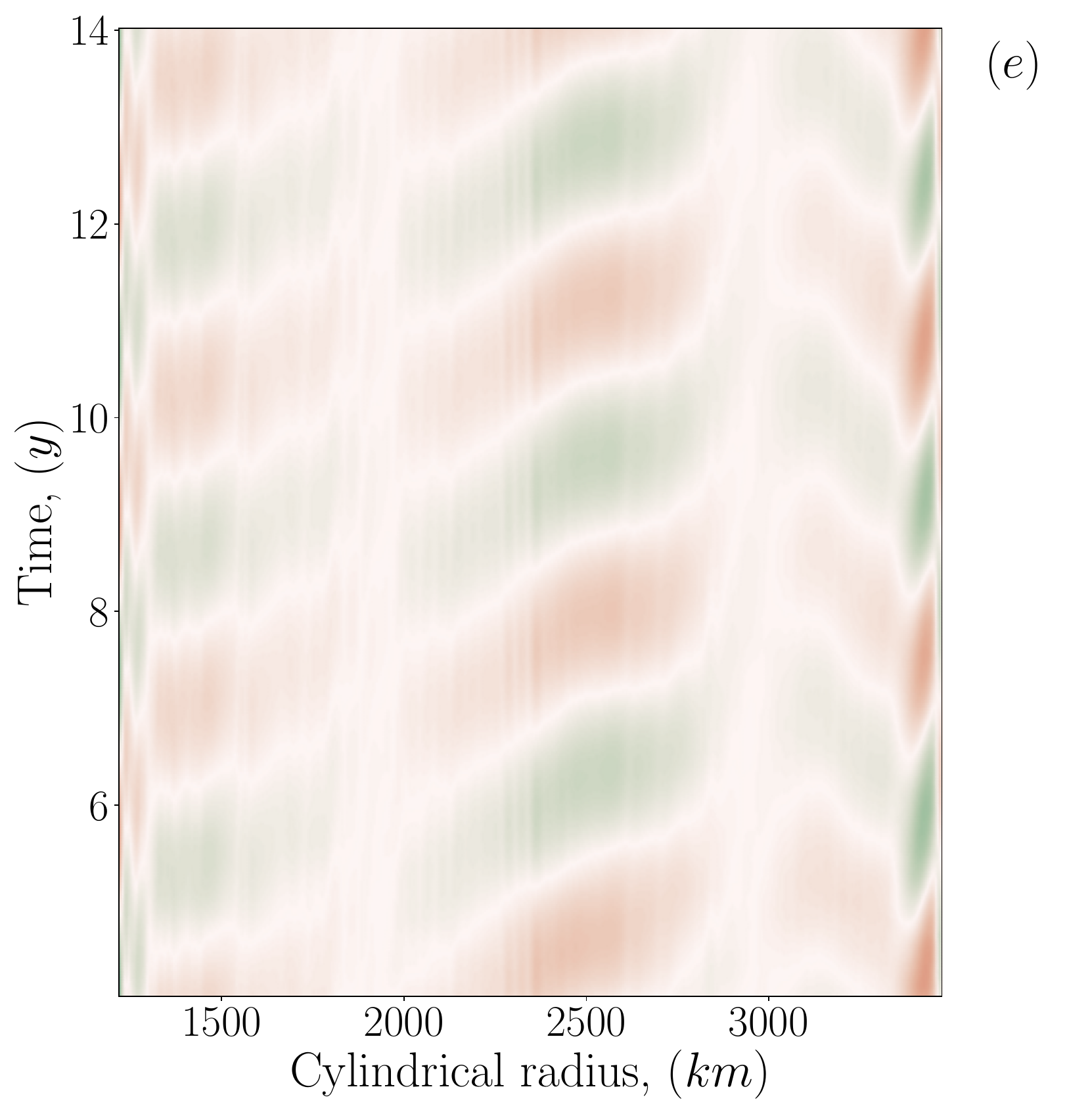}
	\includegraphics[width=0.33\linewidth]{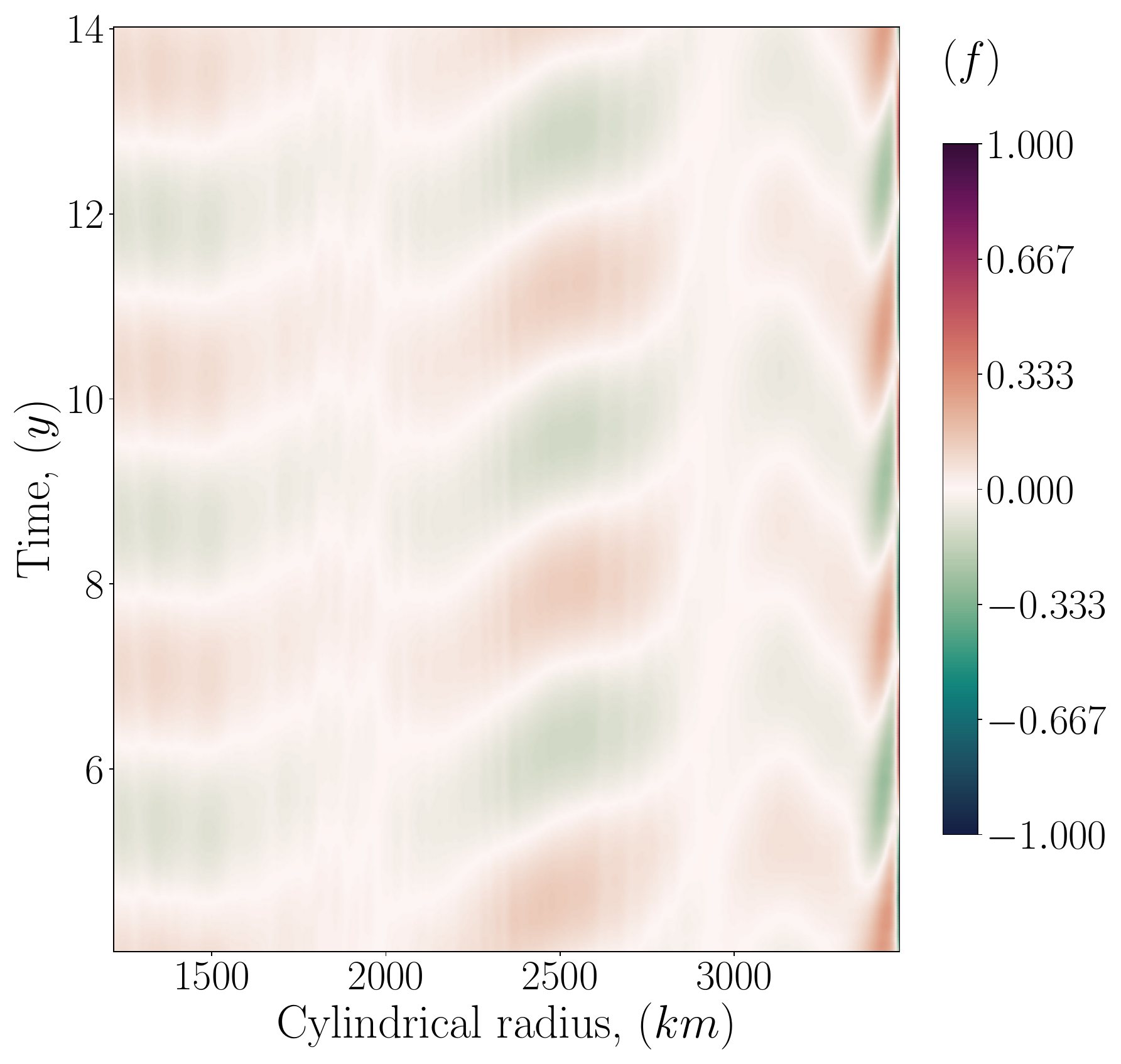}
}
\centering{
    \includegraphics[width=0.31\linewidth]{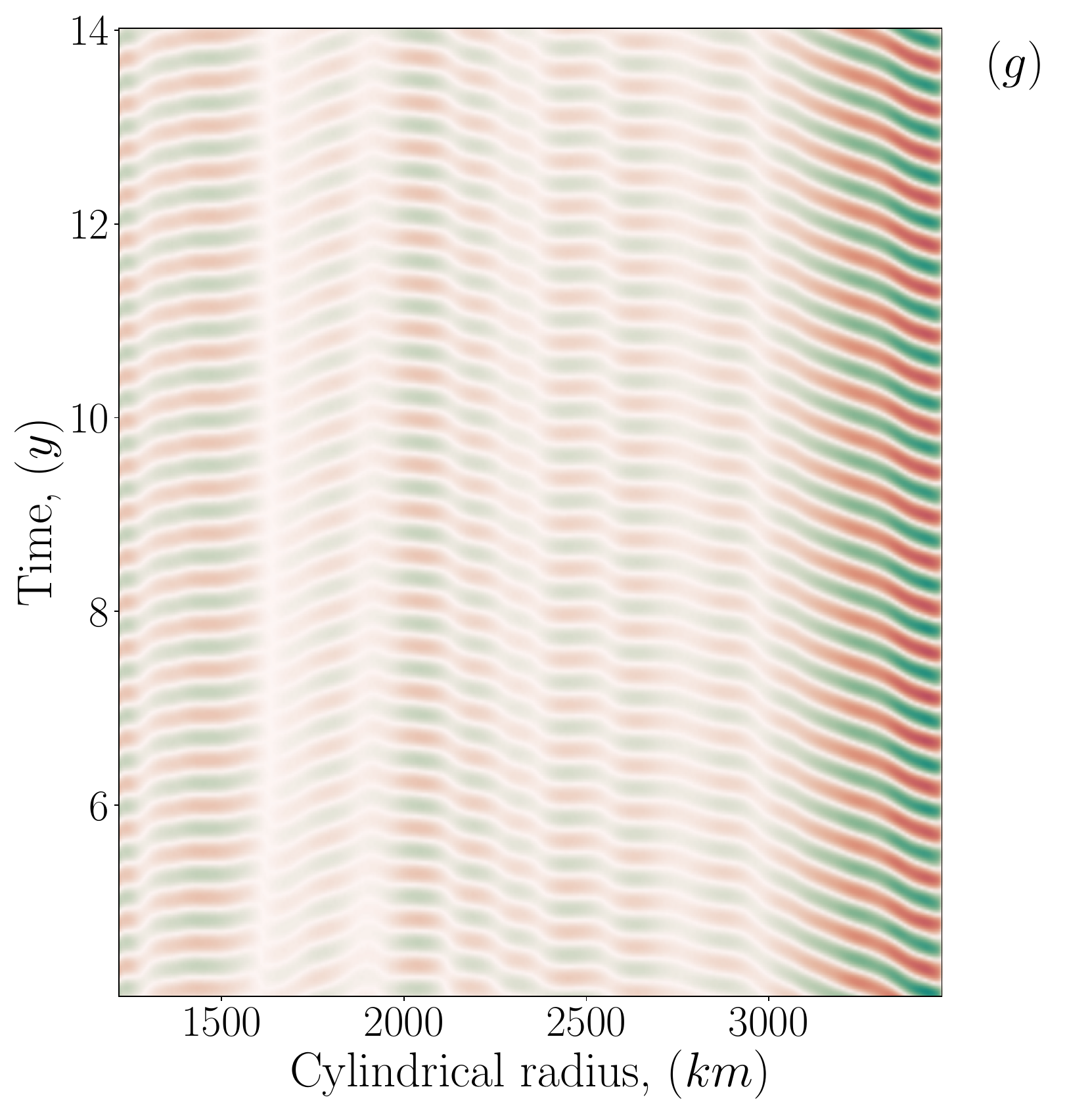}
    \includegraphics[width=0.31\linewidth]{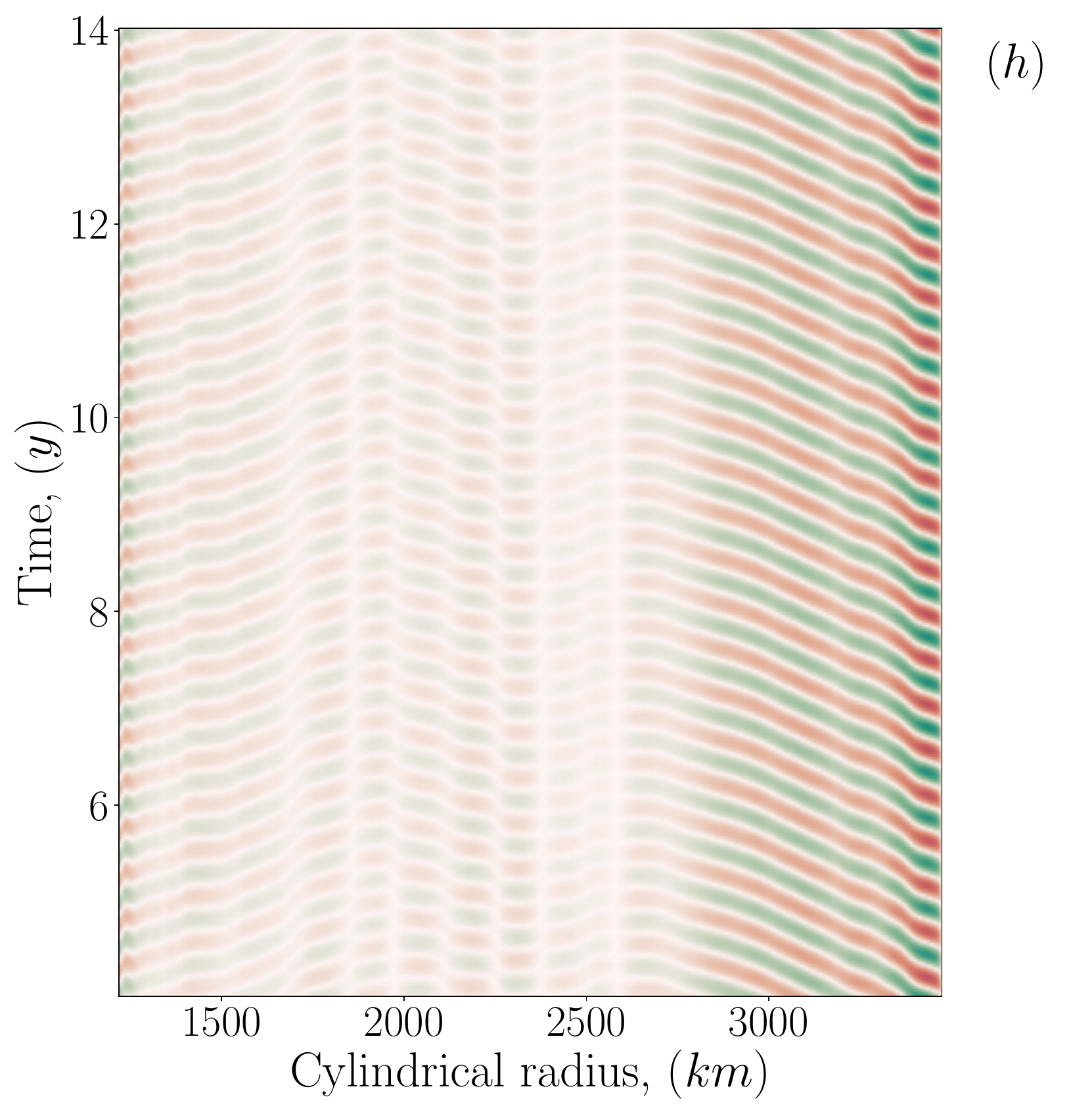}
	\includegraphics[width=0.33\linewidth]{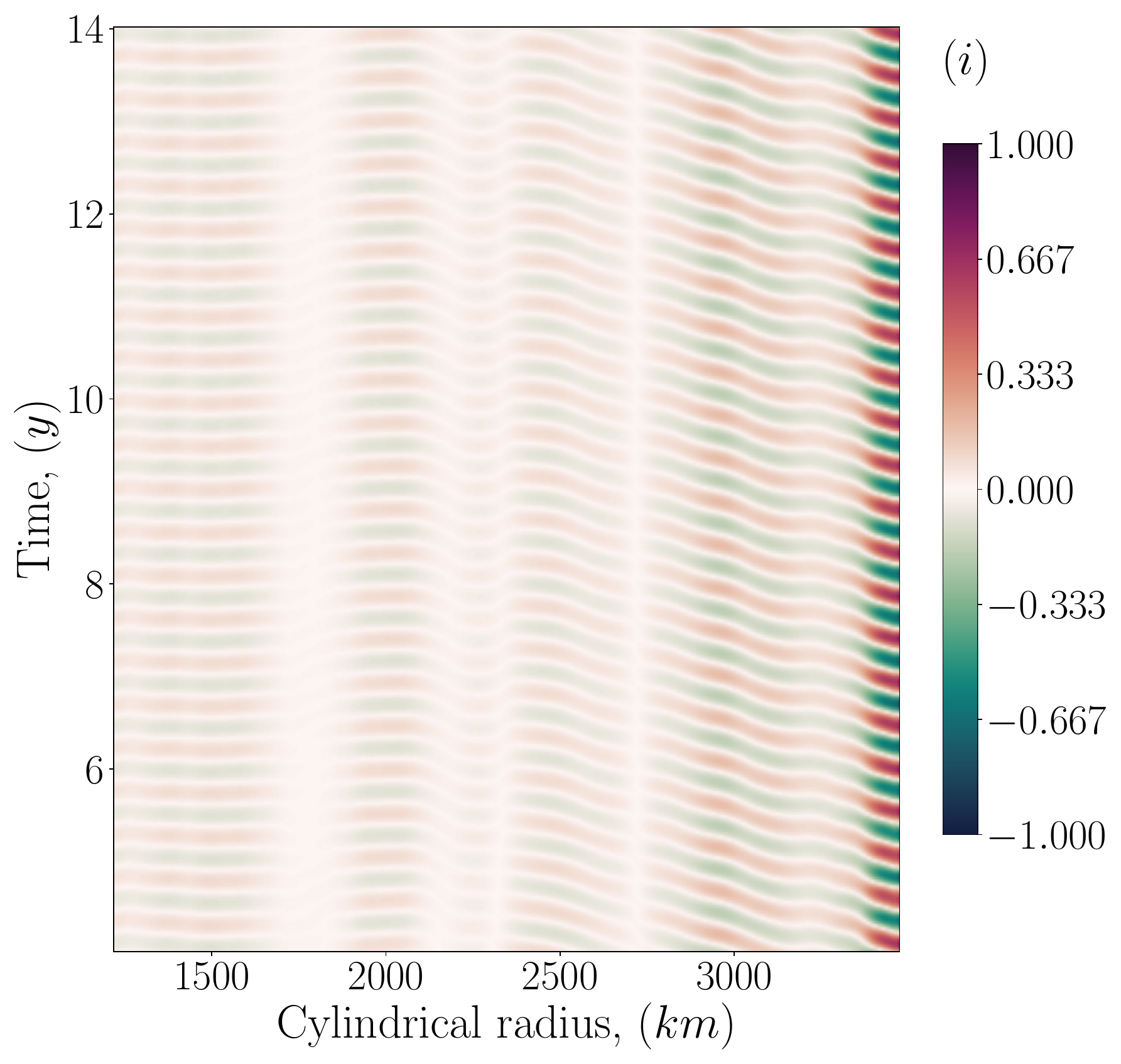}
}
	\caption{
	Temporal evolution of the $z$-averaged and curled inertia 
    (left column), Lorentz force 
    (middle column) and Coriolis force 
    (right column) for Case SB-1 using an input period corresponding to $T_i = 4.4\,y$ (panels a--c), 
    for Case CB-1 also using an input period corresponding to $T_i = 4.4\,y$ (panels d--f), 
    and for Case CB-2 using an input period corresponding to $T_i = 0.6\,y$ (panels g--i). 
    The force balance is taken at a particular 'fast' longitude corresponding to where $B_{0,s}$ is the densest.
    Note that all forces have been normalised by the maximum value of the Coriolis force for each case.
    }
	\label{fig:Col_Forces_Sin1S-PB}
\end{figure*}

In Figure~\ref{fig:Col_Forces_Sin1S-PB}, we show three time-cylindrical radius diagrams of axially-averaged and curled forces that sustain the waves as they travel through the outer core.

With a forcing period of $T_i = 4.4\,y$ and regardless of the background magnetic field, we immediately see that inertia does not appear in Cases SB-1 and CB-1 and that only the Lorentz and the Coriolis forces are involved in the force balance, unequivocally characterising QG-MC waves (Fig.~\ref{fig:Col_Forces_Sin1S-PB}~a--f).
This is consistent with what has already been reported in our previous study, using a different initial setup \citep{barrois2024characterization}, but here the torsional Alfvén waves are remarkably absent due to the non-axisymmetric nature of the forcing in this study.
It is clear that QG-MC waves arise at the top of the core with a defined temporal and spatial wavelength, even if the travel of the waves in the bulk is more complex and disturbed when using a realistic background state (Fig.~\ref{fig:Col_Forces_Sin1S-PB}~d--f).
Note that in all cases, the observed period of the QG-MC waves at the top of the core remains close to that of the input period.

At a faster input period $T_i = 0.6\,y$ in the complex background state, we can observe that all three quantities are involved in Case CB-2 and that a balance between inertia and the Lorentz force -- characterising QG-Alfvén waves -- dominates the signals (Fig.~\ref{fig:Col_Forces_Sin1S-PB}~g--i). 
QG-MC waves remain present in the simulation -- as revealed by the subsisting balance between the Lorentz and the Coriolis forces -- but they seem to propagate inward similarly to the QG-Alfvén waves and we do not clearly observe outward propagating QG-MC waves.
In fact, at this higher frequency, QG-MC waves begin to be superseded by inward QG-Alfvén waves and the latter will be retrieved at the core surface in the components of the fields as it has already been observed in Fig.~\ref{fig:Maps-CMB_t_evolution_PB_faster}~(a--d).

The results from Fig.~\ref{fig:Maps-CMB_t_evolution_PB_faster} and Fig.~\ref{fig:Col_Forces_Sin1S-PB} clearly indicate that the input frequency of the initial perturbation is instrumental in controlling the resulting waves that can be recovered at the core mantle boundary.
Therefore, we conclude that there is a constrained range of possible forcing frequencies for the QG-MC waves to be detectable at the core surface.
This range varies with the background magnetic field intensity but does not change dramatically with the background state complexity.

\subsection{Dispersion relation}
\label{sec:disp_rel}

\cite{gillet2022satellite} have derived a dispersion relation -- correlating the pulsation of a wave $\omega$ to its cylindrical radial wavenumber $k_s$ --  for QG-Alfvén and QG-MC waves which can be written
\begin{align}
\label{eq:disp_rel_QGAMC_k0}
\omega = v_{\cal A} k_s \left( \left( \dfrac{k_0}{k_s} \right)^3 \pm \sqrt{1 + \left( \dfrac{k_0}{k_s} \right)^6} \right)\,,
\end{align}
making use of $k_0$, a radial wavenumber above which only Alfvén waves exist, and below which QG-MC waves can be separated from inertial waves, reading
\begin{align}
\label{eq:disp_rel_k0}
k_0 = \left( \dfrac{m \Omega}{v_{\cal A} h^2}  \right)^{1/3}\,,
\end{align}
which implies that the transition between the QG-MC and the QG-Alfv\'en waves is controlled by the base state because this number $k_0$ depends on the azimuthal wavenumber $m$ and on the Alfvén velocity $v_{\cal A}$, which itself depends on the strength of the equatorial background magnetic field $B_{0,s}$, as
\begin{align}
\label{eq:Va}
v_{\cal A}(s, \phi) \equiv \displaystyle\sqrt{\dfrac{1}{2h\,\rho\mu}\int_{-h}^{h} B_{0,s}^2\, \mathrm{d}z}\,.
\end{align}
In the above expressions, $h \equiv \sqrt{s_\mathrm{CMB}^2 - s^2}$ is the half-height of a cylinder aligned with the rotation axis at a cylindrical radius $s$.

In the limit of small wavelengths ($k_s \gg k_0$), we recover the first order of the dispersion relation and the waves frequencies only weakly depart from the Alfvén wave frequency, {\it i.e.}
\begin{align}
\label{eq:disp_rel_QGA}
\omega_{\cal A} \simeq \pm v_{\cal A} k_s + \dfrac{m \Omega}{k_s^2 h^2}\,,
\end{align}
while in the opposite limit ($k_s \ll k_0$), we can separate the inertial Rossby waves and the QG-MC waves, with the latter's frequency being
\begin{align}
\label{eq:disp_rel_QGMC}
\omega_\mathrm{MC} = - \dfrac{v_{\cal A}^2 h^2 k_s^4}{2 m \Omega}\,,
\end{align}
\citep{gillet2022satellite}.
Note that Eq.~(\ref{eq:disp_rel_k0}) reveals that the increasing number of symmetries and the increasing strength of the base state have opposing effects on $k_0$ -- because $m$ and $v_{\cal A}$ are respectively at the numerator and the denominator of this expression -- so it is not trivial to anticipate the effect of using a more complex background on the transition between the QG-MC and the QG-Alfvén waves from this dispersion relation.


\begin{figure*}
\centering{
	\includegraphics[width=.99\linewidth]{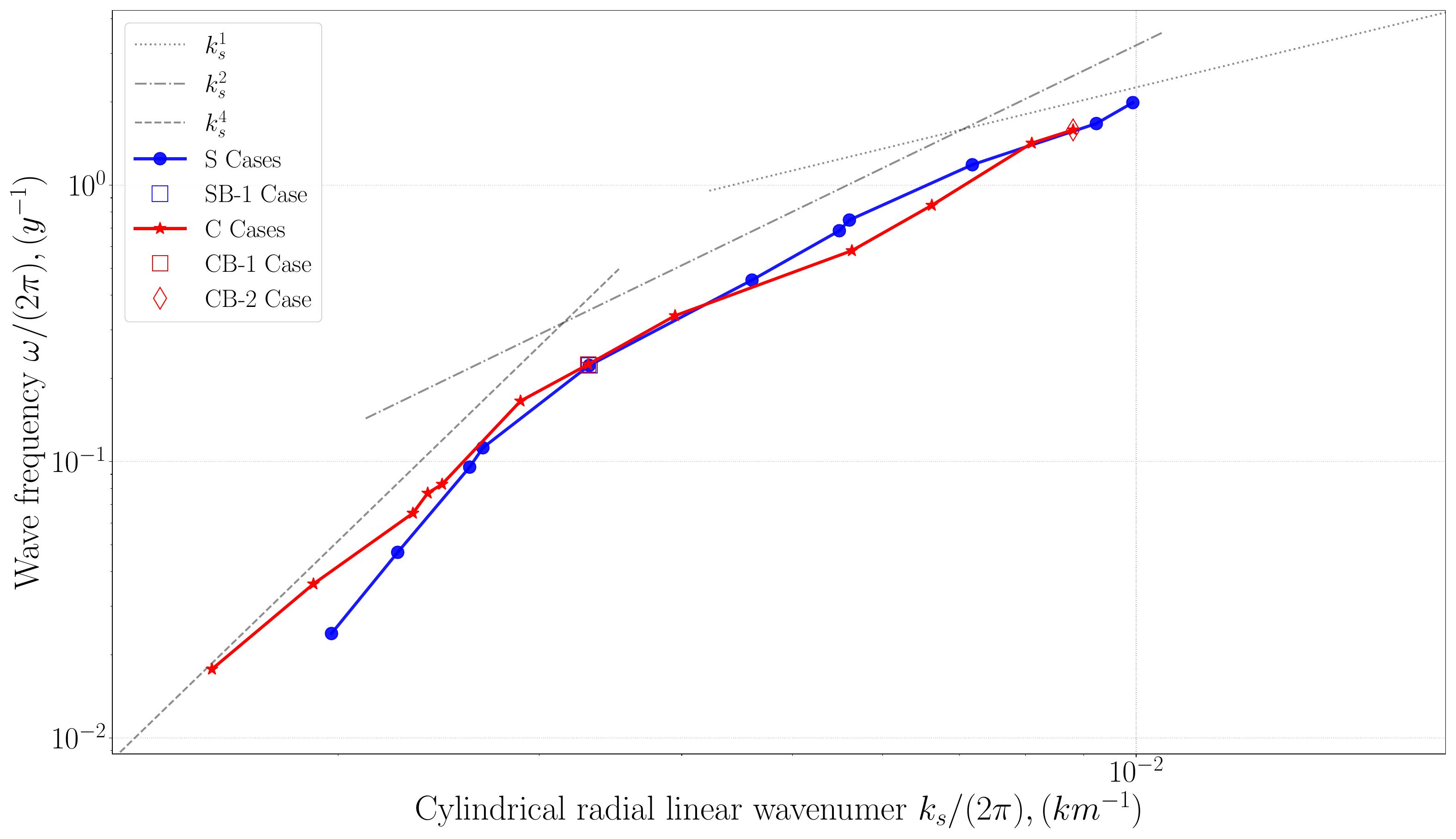}}
	\caption{
    Evolution of the cylindrical radial linear wavenumber $k_s / (2\pi)$ as a function of the frequency $\omega / (2\pi)$, {\it i.e.} observed dispersion relation of the QG-MC -- QG-Alfvén system for several cases computed at high rotation rate and low attenuation 
    using a simple background magnetic field (blue dots), 
    and the complex background magnetic field (red stars).
    Analytical slopes for the dispersion relation are also displayed on this figure.
	}
	\label{fig:Dispersion_Evolution_All}
\end{figure*}

We thus examine the dispersion relation for QG-MC waves in Figure~\ref{fig:Dispersion_Evolution_All} which displays the evolution of the wave's frequency $\omega / 2 \pi$ as a function of its cylindrical radial linear wavenumber $k_s / 2 \pi$ for several series of cases using both the simple 
and the complex background magnetic fields.
We find that the slow frequencies (bottom left of Fig.~\ref{fig:Dispersion_Evolution_All}) are consistent with QG-MC wave properties, {\it i.e.} the dispersion relation Eq.~(\ref{eq:disp_rel_QGMC}) and Fig.~\ref{fig:Maps-CMB_t_evolution_PB},~\ref{fig:Maps-CMB_t_evolution},~\ref{fig:Col_Forces_Sin1S-PB}~a--f.
At higher frequencies (top right of Fig.~\ref{fig:Dispersion_Evolution_All}) both curves flatten which is coherent with the properties of QG-Alfvén waves, {\it i.e.} Eq.~(\ref{eq:disp_rel_QGA}) and Fig.~\ref{fig:Maps-CMB_t_evolution_PB_faster},~\ref{fig:Col_Forces_Sin1S-PB}~g-i.
In addition, we can see that the dispersion relation follows the same evolution in all cases although the 'inflexion point' of each curve is different due to the variation of $k_0$ with the intensity of the background magnetic field.
We can indeed observe that the transition from a $k_s^4$ to a $k_s^1$ slope in Fig.~\ref{fig:Dispersion_Evolution_All} happens between $k_s \approx 40$ and $100$ which is consistent with $k_0$ having a value of $42$ in the C Cases configuration, and a value of $45$ in the S Cases configuration. 
These observations assert that the dispersion relation derived in \cite{gillet2022satellite} is a satisfying approximation.
It also confirms again that the frequency range of the QG-MC waves in the geomagnetic records is restrained.

Thus, our results imply that there is a frequency window for observing the QG-MC waves at the Earth's core surface.
From the complex cases of Fig.~\ref{fig:Dispersion_Evolution_All}, we can estimate that signals with a periodicity much shorter than $\simeq 2.8~y$ will be QG-Alfvén waves. 
Note that we were unable to recover the dispersion relation for periodicities longer than $\simeq 57~y$, mainly because the increasing cylindrical radial wavelengths (decreasing wavenumbers) of the QG-MC waves get longer than the size of the modeled shell. 

\section{Discussion}
\label{sec:Conclusion}

This study confirms that QG-Alfvén and QG-MC waves can be sustained by a variety of non-axisymmetric background magnetic fields.
The non-axisymmetric monochromatic periodic forcing of our model generates QG-MC waves and complex interactions at the core surface produce inward propagating QG-Alfvén waves, 
as previously observed in a different setup \citep{barrois2024characterization}.
The predominance of one rather than the other directly depends on the input frequency of the perturbation in the simulations.
In the limit of large wavelengths ($k_s \ll k_0$) -- corresponding to low input frequencies -- QG-MC waves are recovered at the CMB, and in the opposite limit of small wavelengths ($k_s \gg k_0$) -- corresponding to higher input frequencies -- QG-Alfvén waves 
arise at the top of the core and supersede QG-MC waves.
From the complex background magnetic field cases, we estimate that QG-MC waves with periods longer than $\gtrsim 2.8~y$ can in principle be observable at the top of the Earth core, while waves at the CMB with periods much shorter than $\ll 2.8~y$ are likely QG-Alfvén waves.
Note that in our configuration, waves with periods much longer than $\simeq 57~y$ become hard to recover (which anyway exceed the high resolution geomagnetic temporal series we currently have access to). 

The extent of the frequency range where QG-MC waves can arise at the core surface mainly depends on the strength of the background magnetic field, and our consideration of both simple and complex background magnetic fields shows that geometry does not dramatically change our conclusions (as long as they remain non-axisymmetric \citep{gerick2024interannual}).
This is encouraging for geomagnetic data assimilation \cite[{\it e.g.},][]{lesur2022rapid,istas2023transient} and for predicting the magnetic field evolution such as with the International Geomagnetic Reference Field \citep[IGRF,][]{alken2021international} because the robustness of our results against changes in the background state holds the promise of increasing the predictive power of models based on QG-MC wave patterns analysis.
However, because the CMB signature of QG-MC waves is only marginally affected by the base state, concerns can be raised for capturing insights on the deep structures of the magnetic field in the outer core as hoped by {\it e.g.} \cite{gillet2022satellite}.

The patterns arising in the radial component of the magnetic field at the core surface indicate that the QG-MC waves are nevertheless sensitive to the background magnetic field morphology near the core surface.
The QG-MC waves probably retain probing capabilities as they are channeled by the dense regions of $B_{0,s}$ and we have seen in the geophysical complex magnetic background cases that regional heterogeneities reminiscent of recently observed geomagnetic jerks are retrieved at the core surface.
These results suggest that the QG-MC waves at least provide insights on the magnetic field near the core surface and that the information they might carry on the deep state of the magnetic field remains to be refined.

As a final note, the study of waves to probe the Earth's deep layers \citep[see {\it e.g.},][]{gillet2022dynamical,triana2022core,schwaiger2024wave}, and the use of reduced models are applicable to a variety of systems and could also bring knowledge to other celestial bodies from which geophysical signals are currently recorded by satellites in the framework of spacecraft missions such as, for example, the JUNO mission \citep{bolton2010juno}. 

\section*{Acknowledgements}
O.Barrois has received funding from the European Research Council (ERC) {\bf GRACEFUL} Synergy Grant No. 855677.
We are grateful to Mioara Mandea for fruitful discussions within the course of this project.
Numerical computations were performed on the S-CAPAD platform, IPGP, France.

\section*{Data availability}

Additional data are provided and available at: \hlit{look at dataverse.ipgp to store data}. 
The python package {\tt parobpy} and the scripts used to produce the results shown in this manuscript are available at \url{https://github.com/OBarrois/parobpy}.

\bibliographystyle{gji}

\appendix
\section{Methods details}
\label{sec:Append-A-Methods}

\subsection{Linearised Magneto-hydrodynamic equations}
\label{sec:MHD-lin_equations}

Our reduced system of linearised equations solves for the velocity perturbation ${\bm u}$ and magnetic perturbation ${\bf b}$ fields, with ${\bf U} = {\bf U}_0 + {\bm u}$ -- note that the amplitude of the background velocity field $U_0$ is arbitrary -- and ${\bf B} = {\bf B}_0 + {\bf b}$.
It follows that the temporal evolution of the perturbation fields in our models reads
\begin{align}
\label{eq:momentum_noT_linearised}
\dfrac{\partial {\bm u}}{\partial t} + \dfrac{2}{\lambda}\,{\bm e}_z \times {\bm u} = - \nabla p + \dfrac{1}{Pm\,\lambda}\,\left[ (\nabla \times {\bf b}) \times {\bf B}_0 + (\nabla \times {\bf B}_0) \times {\bf b} \right] + \dfrac{Pm}{S}\,\nabla^2 {\bm u}\,,
\end{align}
\begin{align}
\label{eq:induction_linearised}
\dfrac{\partial {\bf b}}{\partial t} = \nabla \times ({\bm u} \times {\bf B}_0) + \dfrac{1}{S}\,\nabla^2 {\bf b}\,. 
\end{align}
Where the perturbations ${\bm u}$ and ${\bf b}$ are triggered by an initial perturbation in the force balance (see~\ref{sec:Methods}\ref{sec:Init}).

Using the Alfvén timescale $\tau_{\cal A} = d\,\sqrt{\rho\,\mu}/B_0$ as the reference for time -- where $B_0$ is the rms value of the background magnetic field --, the shell thickness $d$ as the reference for length, and the Elsasser unit $\sqrt{\Omega\,\eta\,\rho\,\mu}$ as the reference for magnetic field strength -- where $\eta$ is the magnetic diffusivity of the fluid --, our system is controlled by the dimensionless Lehnert $\lambda$, Lundquist $S$, and magnetic Prandtl $Pm$ numbers -- whereas the Ekman number $Ek$ is included as a reference --, respectively defined as
\begin{align}
\label{eq:adim_par}
\lambda = \dfrac{B_0}{\Omega d \sqrt{\rho\,\mu}}\,, \; S = \dfrac{d B_0}{\eta \sqrt{\rho\,\mu}}\,, \; Pm = \dfrac{\nu}{\eta}\,, \; Ek = \dfrac{\nu}{\Omega d^2}\,,
\end{align}
for which their Earth's core estimates are approximately $\lambda \sim {\cal O}(10^{-4})$, $S \sim {\cal O}(10^{5})$, $Pm \sim {\cal O}(10^{-6})$ and $Ek \sim {\cal O}(10^{-15})$ \citep{de1998viscosity, gillet2010fast, pozzo2014thermal}, and which can also be expressed as timescales, such that
\begin{align}
\label{eq:adim_par_tau}
\lambda = \dfrac{\tau_\Omega}{\tau_{\cal A}}\,, \; S = \dfrac{\tau_\eta}{\tau_{\cal A}}\,, \; Pm = \dfrac{\tau_\eta}{\tau_\nu}\,, \; Ek = \dfrac{\tau_\Omega}{\tau_\nu}\,,
\end{align}
where $\nu$ is the kinematic viscosity of the fluid, $\tau_\eta = d^2/\eta$ is the magnetic diffusive timescale, $\tau_\nu = d^2/\nu$ is the viscous diffusive timescale, and $\tau_\Omega = 1/\Omega$ is the rotation timescale.

Introducing the Elsasser number $\Lambda$, a dimensionless measure of the background magnetic field strength, that is
\begin{align}
\label{eq:Elsasser}
\Lambda = \dfrac{B_0^2}{\Omega\,\eta\,\rho\,\mu} = \dfrac{S^2\,Ek}{Pm} = \dfrac{Ek}{\lambda^2\,Pm}\,,
\end{align}
the simple background field has a dimensionless strength of $\Lambda = 1.77$ and the complex background field has a dimensionless strength of $\Lambda = 3.13$.
In both series, we vary the input pulsation of the perturbation from $\omega_i = 0.2$ up to $\omega_i = 25$ (where times are expressed in terms of the Alfvén timescale $\tau_{\cal A}$).

\subsection{Numerical implementation}
\label{sec:Numerics}

The numerical implementation of our model is based on a second-order finite-difference scheme up to a maximum of $N_r$ grid points in the radial direction, and on a spherical harmonic decomposition of the velocity field  ${\bf U}$ and the magnetic ${\bf B}$ field up to a maximum degree and order $\ell_\text{max} = m_\text{max}$ in the horizontal direction.
The spherical harmonic transforms are handled using \texttt{SHTns}\footnote{\url{https://bitbucket.org/nschaeff/shtns}} \citep{schaeffer2013efficient}.
The Message Passing Interface ({\tt MPI}) library is used for the parallelisation of the code.
And to time-step the equations of our system, a second-order, semi-implicit scheme is used.

The solutions of our problem are approximated making use of the hyperdiffusivity of the small length-scales of the velocity field, following \cite{nataf2015turbulence,aubert2017spherical,aubert2019approaching}, such that
\begin{align}
\label{eq:hdif-vel}
\nu_\mathrm{eff} =  \nu\,q_H^{\ell-\ell_H}\, \; \mathrm{for }\, \; \ell \geq \ell_H\,,
\end{align}
where $\ell_H$ is the cut-off degree above which the hyperdiffusion smoothly increases and affects the small length-scales of the velocity field -- the magnetic field is never affected in our model --, and $q_H$ is the parameter controlling the increasing level of the hyperdiffusivity.
Values for $\ell_H$ and $q_H$ have been chosen such that the solutions show a satisfying convergence of the kinetic and magnetic energy spectra.
To test the convergence of our solutions, the hyperdiffusion parameters have been varied in a series of different cases and we have not observed any major changes in the average properties \citep[see our previous study][for more details]{barrois2024characterization}.
Note that in this study, we have used a fixed strength $q_H = 1.05$, a fixed cut-off harmonic degree $l_H = 30$ for the hyperdiffusion, and a fixed grid-size of $(N_r, \ell_{\text{max}}) = (450, 133)$.

\onecolumn
\section{Results of numerical simulations}
\label{sec:Append-B-Results}

\begin{longtable}{p{0.07\textwidth} p{0.07\textwidth} p{0.14\textwidth} p{0.07\textwidth} p{0.09\textwidth} p{0.09\textwidth} p{0.07\textwidth} p{0.20\textwidth}}
\caption{Summary of the numerical simulations computed in this study using $r_i/r_o = 0.35$ as suited for the Earth's core.
In the first column, the letter S denotes the runs using a simple background magnetic field, the letter C denotes the runs using a geophysical complex background magnetic field, while codes such as SB-1 denote specific cases shown in the main text.
$S$ is the Lundquist number, $\lambda$ is the Lehnert number, $\Lambda$ is the Elsasser number of the background magnetic field, $\omega_i$ is the input pulsation of the perturbation, $\omega_o$ is the pulsation of the output waves at the core surface, and $k_{s, o}$ is the output cylindrical radial wavenumber of the waves at the core surface.
The last column reports which type of wave is observed at the core surface.
All runs have been computed using an Ekman number $Ek = 1 \times 10^{-7}$, a magnetic Prandtl number $Pm = 0.144$, a fixed strength $q_H = 1.05$ and a fixed cut-off harmonic degree $l_H = 30$ for the hyperdiffusion, and a grid-size of $(N_r, \ell_{\text{max}}) = (450, 133)$.
}
\label{tab:run_list} \\
\hline
Case & $S$ & $\lambda$ & $\Lambda$ & $\omega_i$ & $\omega_o$ & $k_{s, o}$ & Waves at CMB \\
\hline
\endfirsthead

\hline
Case & $S$ & $\lambda$ & $\Lambda$ & $\omega_i$ & $\omega_o$ & $k_{s, o}$ & Waves at CMB \\
\hline
\endhead

\hline
\multicolumn{7}{c}{Continued on next page $\ldots$} \\
\hline
\endfoot

\hline
\hline
\endlastfoot

S & $1596$ & $1.11 \times 10^{-3}$ & $1.77$ & $0.292$ & $0.3$ & $28$ & QG-MC \\
S & $1596$ & $1.11 \times 10^{-3}$ & $1.77$ & $0.585$ & $0.59$ & $32$ & QG-MC \\
S & $1596$ & $1.11 \times 10^{-3}$ & $1.77$ & $1.170$ & $1.2$ & $37$ & QG-MC \\
S & $1596$ & $1.11 \times 10^{-3}$ & $1.77$ & $1.428$ & $1.41$ & $38$ & QG-MC \\
SB-1 & $1596$ & $1.11 \times 10^{-3}$ & $1.77$ & $2.856$ & $2.8$ & $47$ & QG-MC \\
S & $1596$ & $1.11 \times 10^{-3}$ & $1.77$ & $5.712$ & $5.7$ & $65$ & QG-MC \\
S & $1596$ & $1.11 \times 10^{-3}$ & $1.77$ & $9.359$ & $9.4$ & $80$ & QG-MC \\
S & $1596$ & $1.11 \times 10^{-3}$ & $1.77$ & $14.28$ & $14.9$ & $102$ & QG-MC--Alfvén\\
S & $1596$ & $1.11 \times 10^{-3}$ & $1.77$ & $19.99$ & $21.0$ & $131$ & QG-Alfvén \\
S & $1596$ & $1.11 \times 10^{-3}$ & $1.77$ & $24.28$ & $25.0$ & $141$ & QG-Alfvén \\
C & $2124$ & $1.48 \times 10^{-3}$ & $3.13$ & $0.217$ & $0.22$ & $22$ & QG-MC \\
C & $2124$ & $1.48 \times 10^{-3}$ & $3.13$ & $0.435$ & $0.45$ & $27$ & QG-MC \\
C & $2124$ & $1.48 \times 10^{-3}$ & $3.13$ & $0.869$ & $0.82$ & $33$ & QG-MC \\
C & $2124$ & $1.48 \times 10^{-3}$ & $3.13$ & $1.061$ & $1.04$ & $35$ & QG-MC \\
C & $2124$ & $1.48 \times 10^{-3}$ & $3.13$ & $1.170$ & $1.00$ & $34$ & QG-MC \\
C & $2124$ & $1.48 \times 10^{-3}$ & $3.13$ & $2.123$ & $2.08$ & $41$ & QG-MC \\
CB-1 & $2124$ & $1.48 \times 10^{-3}$ & $3.13$ & $2.856$ & $2.85$ & $47$ & QG-MC \\
C & $2124$ & $1.48 \times 10^{-3}$ & $3.13$ & $4.245$ & $4.24$ & $56$ & QG-MC \\
C & $2124$ & $1.48 \times 10^{-3}$ & $3.13$ & $7.429$ & $7.28$ & $80$ & QG-MC \\
C & $2124$ & $1.48 \times 10^{-3}$ & $3.13$ & $10.61$ & $10.6$ & $94$ & QG-MC \\
C & $2124$ & $1.48 \times 10^{-3}$ & $3.13$ & $18.04$ & $17.8$ & $115$ & QG-Alfvén \\
CB-2 & $2124$ & $1.48 \times 10^{-3}$ & $3.13$ & $19.99$ & $19.9$ & $125$ & QG-Alfvén \\
\end{longtable}

\label{lastpage}

\end{document}